\documentclass[prb,aps,twocolumn]{revtex4}
\usepackage[dvips]{graphicx}
\usepackage{latexsym}
\usepackage{amsmath,enumitem}

\newcommand{\vect}[1]{\boldsymbol{#1}}

\renewcommand{\(}{\left(}
\renewcommand{\)}{\right)}
\renewcommand{\[}{\left[}
\renewcommand{\]}{\right]}

\begin{document}

%\preprint{cond-mat/0305xxx}

\title{Stripe melting and quantum criticality in correlated metals}
\author{David F. Mross, and T. Senthil}
\affiliation{Department of Physics, Massachusetts Institute of Technology,
Cambridge, MA 02139, USA }

\date{\today}
\begin{abstract}
We study theoretically quantum melting transitions of stripe order in a metallic environment, and the associated reconstruction of the electronic Fermi surface. We show that such quantum phase transitions can be continuous in situations where the stripe melting occurs by proliferating pairs of dislocations in the stripe order parameter without proliferating single dislocations. We develop an intuitive picture of such phases as ``Stripe Loop Metals" where
the fluctuating stripes form closed loops of arbitrary size at long distances.
We obtain a controlled critical theory of a few different continuous quantum melting transitions of stripes in metals . At such a (deconfined) critical point the fluctuations of the stripe order parameter are strongly coupled, yet tractable. They also decouple dynamically from the Fermi-surface. We calculate many universal properties of these quantum critical points. In particular we find that the full Fermi-surface and the associated Landau quasiparticles remain sharply defined at the critical point. We discuss the phenomenon of Fermi surface reconstruction across this transition and the effect of quantum critical stripe fluctuations on the superconducting instability. We study possible relevance of our results to several phenomena in the cuprates.

\end{abstract}
\newcommand{\be}{\begin{equation}}
\newcommand{\ee}{\end{equation}}
\maketitle
\section{Introduction}
In the last several years evidence for the occurrence of stripe and related orders has accumulated in many underdoped cuprates. Static charge and spin stripes have long been known to occur in some La-based cuprates\cite{review}. Other cuprates have been thought to have incipient stripe ordering (``dynamic stripes") which can be pinned locally around impurities or near vortex cores in the low temperature superconducting state. Such pinning of incipient stripe order is routinely observed in STM experiments in the Bi-based\cite{jenny} and oxychloride\cite{hanaguri} cuprates. Further a number of phenomena in the YBCO family have indicated the possible role of incipient stripe order. Quantum oscillations\cite{leyraud} seen in high magnetic field and low temperature have been interpreted\cite{norman} in terms of the reconstruction of a large Fermi surface by stripe order. Such a Fermi surface reconstruction might also explain the sign of the Hall effect\cite{leboeuf} and various other transport properties of underdoped YBCO at low temperature and in high magnetic fields\cite{striperec}. Several experiments also show that YBCO undergoes a rapid crossover to a regime of enhanced orthorhombicity at a temperature that roughly tracks the pseudogap temperature\cite{ando,keimer,daou}. This is reasonably associated with enhanced electronic nematic order which is naturally a precursor of stripe order. Most recently direct evidence for charge stripes in YBCO has been obtained in an NMR experiment in high magnetic fields\cite{julien}.

The stripe ordering seems to disappear with increasing doping. The Fermi surface of overdoped Tl-2201 has been mapped out in great detail through angle dependent magnetoresistance studies\cite{hussey1}, quantum oscillations\cite{hussey2}, and angle resolved photoemission experiments\cite{damascelli}. All these probes clearly and convincingly demonstrate the existence of a large band structure-like Fermi surface at low temperature. The absence of any Fermi surface reconstruction strongly suggests the absence of stripe ordering.

These developments sketch a picture of the evolution of the low temperature ``underlying normal" state electronic properties upon moving from the overdoped to the underdoped side. Decreasing doping tends to produce stripe ordering which reconstructs the large Fermi surface. It is natural then to expect the existence of a quantum phase transition between the large Fermi surface metal and a stripe ordered metal with reconstructed small Fermi pockets. Such a transition, if second order, might conceivably play a role in the mysterious phenomena characterizing the strange metal regime of optimally doped cuprates. This viewpoint is advocated e.g. by Taillefer in Ref. \onlinecite{taillefer}.

Despite this strong motivation, there is very little theoretical understanding of such quantum phase transitions. In a weakly interacting Fermi liquid the stripe order may be viewed simply as a uni-directional charge density wave (CDW) or spin density wave (SDW). The transition to this kind of order may then be described in terms of a fluctuating order parameter coupled to the particle-hole continuum of the metallic Fermi surface. Such a theory was formulated by Hertz\cite{hertz,millis} and others in the 1970s and has received enormous attention over the years. Despite this the theory is very poorly understood in two space dimensions (the case relevant to cuprates). Very interesting recent work shows that the low energy physics involves strong coupling between the various gapless degrees of freedom - the resulting theory currently has no controlled description and remains to be understood\cite{abanov1,abanov2,maxsdw}.

In light of this, and in light of the fact that the cuprates are in any case unlikely to be correctly described as weakly interacting Fermi liquids, it is natural to explore alternate strong coupling approaches to phase transitions associated with stripe ordering\cite{electronlc} and the associated Fermi surface reconstruction. Specifically it is interesting to view this transition as a melting of stripe order by quantum fluctuations. Could the stripes melt through a continuous quantum phase transition? What is the nature of the resultant melted phase? How does the metallic Fermi surface affect (and is affected by) such a putative continuous stripe melting transition? What is the correct description of the universal singularities associated with such a stripe melting quantum critical point?

In a recent short paper we initiated a study of these questions with the main goal of explaining some old neutron scattering experiments in the cuprates. In this paper we expand in detail the ideas on stripe melting we discussed in our earlier work\cite{short}. We provide a few concrete and tractable examples of continuous stripe melting transitions in the presence of a metallic Fermi surface. As expected the stripe melting is accompanied by a reconstruction of the metallic Fermi surface. Remarkably despite this right at the quantum critical point the critical stripe fluctuations decouple dynamically from the particle-hole continuum of the Fermi surface and are described by a strongly coupled though tractable quantum field theory. In the language of renormalization group theory the coupling of the stripe fluctuations to the Fermi surface is a ``dangerously irrelevant" perturbation - though it is irrelevant at the critical fixed point it is relevant at the stripe ordered fixed point and leads to the Fermi surface reconstruction. One consequence of this dangerous irrelevance is that close to the quantum melting transition, the energy scale associated with the onset of stripe order is parametrically larger than the energy scale at which the Fermi surface reconstructs. We determine the universal critical singularities both of the stripe order parameter and of the electronic excitations at the hot spots on the Fermi surface that are connected to each other by the stripe ordering wavevector.

The stripe melting transitions discussed in this paper and in our earlier work\cite{short} are obtained by proliferating some but not all topological defects of the stripe order parameter. The resulting stripe melted phase retains a memory of the long range stripe ordered phase by possessing gapped excitations associated with the unproliferated topological defects. 
Phases of this kind were proposed by Zaanen\cite{zaa} and co-workers and further explored in Refs. \onlinecite{demler,subir,zaanen}. Despite having the same symmetries, these are \emph{not} regular Fermi liquids, due to the unproliferated defects. A clear distinction between these non-trivial stripe liquid phases and the regular Fermi liquid lies in the topological structure, i.e. the former have ground state degeneracies on non-trivial manifolds. While this difference is completely sharp theoretically, it is very difficult to detect with any conventional experimental probe. In particular these phases have large Fermi surfaces with Landau quasiparticle excitations described within the usual Fermi liquid theory paradigm. Their single-electron properties, transport and low-energy thermodynamics are Fermi-liquid like, thus such phases may have easily been mistaken for Fermi liquids. The topological structure associated with the unproliferated defects leads to a fractionalization of the stripe order parameter - this is extremely difficult to probe in experiments.

% Despite having the same symmetries, it is \emph{not} the regular Fermi liquid due to the gapped single dislocations. A clear distinction between the non-trivial stripe liquid phase and the Fermi liquid lies in the topological structure, i.e. the former has a ground state degeneracy on non-trivial manifolds.

%The corresponding stripe liquid phase was first envisaged by Zaanen\cite{zaa} and co-workers and studied further in Refs. \onlinecite{demler,subir,zaanen}. It is strictly distinct (meaning cannot be smoothly connected to) the usual weakly interacting Fermi liquid. However the distinction is extremely subtle and may easily escape detection by any conventional experimental probe. 

\begin{figure}[ht]
\includegraphics[width=.8\columnwidth]{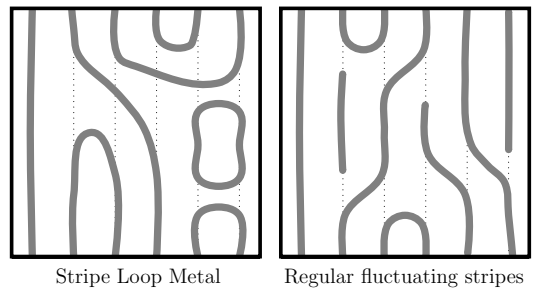}
\caption{Left: In the Stripe Loop Metal phase, only those stripe fluctuations that result in closed loop patterns are allowed. Right: In regular fluctuating stripes, both closed and open patterns are possible.}
\label{fig:slm}
\end{figure}

We provide a simple physical picture of the stripe fluctuations that lead to this kind of stripe melted phase. When the stripes melt the resulting ground state will in general be a quantum superposition of arbitrary stripe configurations. The fluctuating stripe phases described in this paper may be pictured as ones in which the stripes form closed loops of arbitrary size while they fluctuate (see FIG. \ref{fig:slm}). Cutting a stripe loop open to leave an open end for a stripe costs a finite energy and describes an excited state. Such an open end is precisely the gapped unproliferated topological defect that survives the stripe melting transition. This picture is justified and elaborated in detail below. We thus dub such phases ``Stripe Loop Metals". In contrast the conventional Fermi liquid may be viewed as consisting of fluctuating stripes of all kinds including ones with open ends in its ground state.

In the early work such a Stripe Loop Metal was suggested as a candidate for the underdoped cuprates. As it seems extremely unlikely that the ``underlying" normal state in the underdoped cuprates has a large unreconstructed Fermi surface the Stripe Loop Fermi liquid is unlikely to occur in the underdoped side. It is more interesting therefore to explore the possibility that it may describe the ``normal" ground state of the overdoped cuprates. Indeed none of the existing experimental probes of the overdoped normal state are in a position to distinguish between such a stripe-fractionalized Fermi liquid and the conventional Fermi liquid. Further as we demonstrate in this paper the Stripe Loop Fermi liquid
admits a direct and interesting second order transition to the stripe ordered metal with a reconstructed Fermi surface. However our results also demonstrate that the corresponding critical stripe fluctuations cannot by themselves account for most of the observed non-Fermi liquid physics around optimal doping in the cuprates. Nevertheless as we discuss, this kind of stripe melting transition may contain interesting lessons to at least understand the nature of stripe fluctuations near optimal doping. Quantum melting transitions between the stripe ordered metal and the conventional Fermi liquid are more challenging to address, and we leave them for the future.

Apart from developing an intuitive description of the Stripe Loop Metal phase, we also discuss several experimental consequences of the theory of the quantum phase transition to the stripe ordered metal. The availability of a theory of a continuous stripe melting transition enables us to reliably address the phenomenon of Fermi surface reconstruction across
such a transition. For instance we determine how the gap opens at the hot spot as the transition is approached from the stripe melted side and the growth of the quasiparticle weight in the folded
portion of the reconstructed Fermi surface in the stripe ordered phase. We also study the issue of the low temperature superconducting instability of the metal and its interplay with the stripe melting. Not surprisingly we find that the energy scale of the superconducting instability is enhanced as the stripe quantum critical point is approached from either side.

The rest of the paper is organized as follows: In Section \ref{sec:problems} we discuss some of the challenges presented to theory by well known experimental results\cite{AEPPLI}. We then set the stage for discussing phase transitions by first describing various possible stripe order parameters that are of relevance for the cuprates in Section \ref{sec:orderpara}. In Section \ref{sec:phasedia} we discuss possible phase diagrams for stripe melting transitions and the associated defects in the order parameters. The concept of ``Stripe Loop Metals'' is introduced in Section \ref{sec:slm} and connected to the corresponding quantum field theories. In Section \ref{sec:chargemelting} the theories of several charge-stripe melting transitions are presented in detail. Spin-stripe melting transitions are discussed in Section \ref{sec:spin1}. In Section \ref{sec:single} we calculate single particle properties close to the phase transitions. The possibility of pairing by stripe fluctuations within our theory is analyzed in Section \ref{sec:pair}. In Section \ref{sec:exps} we review some results of scaling theory applied to fluctuating stripes. We discuss some existing experimental results as well as predictions of our theory for future experiments. Some more technical details can be found in the appendices.
\section{Challenges from prior experiments}\label{sec:problems}
If quantum criticality associated with onset of stripe order is held responsible for the physics of the strange metal in the cuprates, then it is very important to experimentally establish that critical stripe fluctuations occur in the strange metal regime. There is actually very little information from experiments on critical stripe fluctuations in near optimal cuprates above their superconducting transition termperature.

In an important and well-known experiment, Aeppli et. al. \cite{AEPPLI} measured the dynamic spin-susceptibility of slightly underdoped ($x=0.14$) LSCO near $(\pi,\pi)$ over a wide range of temperatures and wave-vectors. However as emphasized in our previous work\cite{short} the results paint a very intriguing picture of the stripe fluctuations and pose a challenge to theory.
From the scaling of the width of the neutron scattering peak (the inverse correlation length) with temperature they deduced that the dynamical exponent $z\approx 1$. They further found that for low frequencies the peak height scales as $T^{-2}$. Within $z = 1$ scaling, a standard scaling argument shows that this implies an anomalous exponent $\eta \approx 1$ for the critical spin fluctuations (see Sec. \ref{spinscaling}). Such a large value of the anomalous dimension $\eta\approx 1$ of the spin-fluctuations is very unusual for a Landau quantum critical point, but it is quite common for Non-Landau quantum critical points\cite{dccp,chub,kim,motvish,sandvik,tarun,isakov}. Later on in this paper we provide a careful discussion of the data of Ref. \onlinecite{AEPPLI} and highlight the need for further experimental studies.

\begin{figure}[ht]
\includegraphics[width=\columnwidth]{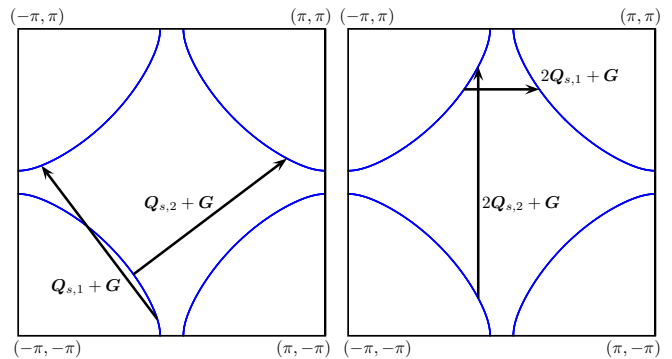}
\caption{The Fermi surface of LSCO at $x=0.15$ as measured in Ref. \onlinecite{arpeslsco} via ARPES, shown in the first Brillouin zone. Left: Hot-spots on the Fermi surface that are connected by the SDW wave vectors. Right: Hot-spots of the CDW wave-vectors.}
\label{fig:vectors}
\end{figure}
\begin{figure}[ht]
\includegraphics[width=3cm]{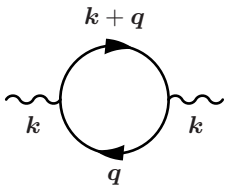}
\caption{Landau damping of the stripe fluctuations is given by the fermionic polarization function near the ordering wave-vector.}
\label{fig:bosonselfenergy}
\end{figure}

We would like to emphasize that in a metal, $z=1$ is very surprising. Since the stripe-ordering vector $\vect Q$ connects two points on the Fermi-surface (see FIG. \ref{fig:vectors}), the stripe-fluctuations should be Landau damped. The propagator of the stripe fluctuations is expected to be modified as (see FIG. \ref{fig:bosonselfenergy})
\begin{align}
 \chi_\text{Stripe}(\vect k,\omega) = &\frac{1}{\omega^2 -v^2(\vect k - \vect Q)^2}\\
&\xrightarrow{\text{\tiny{Landau Damping}}} \frac{1}{\omega^2+i \gamma \omega -v^2(\vect k - \vect Q)^2}\nonumber,
\end{align}
where the damping rate $\gamma$ is determined by the properties of the conduction electrons close to hot spots, i.e. points on the Fermi-surface which are connected by $\vect Q$. For small frequencies $\omega \rightarrow 0$ the quadratic term can be neglected against the damping term and the stripe fluctuations become strongly coupled to the Fermi surface. In particular, we see that $\omega\sim \delta k^2$, where $\delta \vect k = (\vect k - \vect Q)$, i.e. the dynamical exponent $z=2$. Recent careful analyses\cite{abanov1,abanov2,maxsdw} show that at low energies the stripe fluctuations couple even more strongly to the Fermi-surface, and higher order contributions become important. There is currently no controlled description of the resulting theory.

This form of the Landau damping is not unique to the weakly interacting Fermi liquid but rather general in the presence of gapless excitations into which a stripe fluctuation can decay. In Appendix \ref{app:damping} we demonstrate this explicitly for several known non-Fermi liquid metals.

The data of Ref. \onlinecite{AEPPLI} shows dynamical exponent of $z=1$ over all measured temperatures, i.e. no evidence of the expected strong coupling to the Fermi surface is seen. Of course, once a single-particle gap develops, a low-energy stripe fluctuation can no longer decay into particle-hole excitations and effectively decouples from the Fermi-surface. This does not require phase-coherence and is therefore already possible in the pseudogap phase. While such an opening of a gap would explain the observed $z=1$ below the pseudogap temperature $T^*$, it would also predict a dramatic change in the nature of the stripe-correlations upon crossing $T^*$. Above $T^*$ the stripe fluctuations are strongly coupled to and modified by the gapless Fermi surface, while below $T^*$ they decouple. However in the data of Ref. \onlinecite{AEPPLI}, the measured spin-correlations develop smoothly across the pseudogap temperature, which for $x=0.15$ LSCO is around 150 K\cite{zx}.

The apparent message from the experiments is that the critical stripe fluctuations are indifferent to the fate of the electronic Fermi surface. This has also been emphasized by other neutron studies of the cuprates where magnetic scattering near the incommensurate peaks seems to not know about the particle/hole continuum of the Fermi surface at low energies\cite{tranquada}. This state of affairs in experiments should be contrasted with the emerging picture from modern clarifications\cite{abanov1,abanov2,maxsdw} of the standard weak coupling ``Hertz-Millis'' approach to stripe ordering quantum phase transitions: In this theory the stripe fluctuations become more and more strongly coupled to the Fermi surface at low energy.

Given this stark contradiction, we therefore focus our attention on an alternate theory of the stripe ordering transition which describes it as a quantum melting of stripes by proliferation of topological defects.

 \section{Possible stripe order parameters}\label{sec:orderpara}
Throughout the paper we will use the term ``stripe'' as a collective term for various kinds of charge density wave (CDW) and spin density wave (SDW) states. With the cuprates in mind, we focus on two-dimensional systems (a single copper-oxygen plane) with an underlying orthorhombic or tetragonal lattice. The simplest stripe pattern is a
\begin{enumerate}

\item \emph{unidirectional charge stripe}. Here the expectation value of the charge is modulated at some wavevector $\vect Q_c$ (see FIG. \ref{fig:static}), i.e.
\begin{align}
\rho(\vect r) \equiv \langle c^\dagger_{\vect r,\sigma}c_{\vect r,\sigma} \rangle = \rho_0 + \(\rho_{\vect Q_c}e^{i \vect Q_c \cdot \vect r}+ c.c.\),
\end{align}
where $c^\dagger_{\vect r,\sigma}$ creates a spin $\sigma$ electron at $\vect r$ and summation over spin indices is implied. Such unidirectional stripe patters occur naturally in orthorhombic crystals, where there is a preferred direction for $\vect Q_c$. In tetragonal crystals unidirectional order is possible by spontaneously breaking the lattice rotation symmetry (in addition to the lattice translation symmetry along $\vect Q_c$). This case is frequently referred to as \emph{smectic} in the literature.

 When the ordering vector $\vect Q_c$ is commensurate with the lattice, i.e. $\vect Q_c = \frac{2\pi}{ a}\left(m_x,m_y\right)$, where $m_{x,y}$ are integers, this is referred to as a phase with \emph{commensurate charge stripes}. For the cuprates, commensurate stripes with $m_y(m_x)=0$ and a period of $m_x(m_y)=4$ are most commonly observed experimentally.
\item In a \emph{unidirectional spin stripe} the expectation value of the spin undergoes spatial modulations, i.e.
\begin{align}
 \vec S(\vect r) \equiv\frac{1}{2}\langle c^\dagger_{\vect r,\sigma}\vec \tau_{\sigma \sigma'} c_{\vect r,\sigma'}\rangle =e^{i \vect Q_s \cdot \vect r} \vec M+ e^{-i \vect Q_s \cdot \vect r} \vec M^*.
\end{align}
For \emph{collinear} spin order, i.e. $\vec S(\vect r) \times \vec S(\vect r')=0$, it follows that
\begin{align}
&\vec M =e^{i \theta}\vec N,
\end{align}
where $\vec N$ is a real vector. A further possibility is \emph{spiral} order, i.e.
\begin{align}
 \vec M = \vec n_1+ i\vec n_2
\end{align}
where $\vec n_1$ and $\vec n_2$ are real vectors with $\vec n_1 \cdot \vec n_2=0$. With the cuprates in mind, we consider collinear spin-stripes exclusively.
In the cases of interest here, SDW order at $\vect Q_s$ will be accompanied by CDW order at $\vect Q_c =2 \vect Q_s$. In a weak coupling, Landau-Ginzburg approach, this follows since the charge density has the same symmetries as the \emph{square} of the spin density. Since the ordering wave-vectors of charge and spin are tied together, it is sufficient to name either one. We will adopt the convention that the period of a commensurate stripe refers to the charge sector, i.e. a ``period-$m$ stripe'' has period $m$ for the charge, but period $2m$ for spin (see FIG. \ref{fig:neel}).
 \begin{figure}[ht]
\includegraphics[width=\columnwidth]{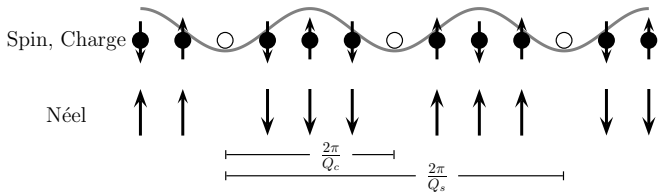}
\caption{A period-4 charge stripe accompanied by an \emph{antiphase} spin stripe. The sign of the N\'eel-vector changes from one charge-stripe to the next.}
\label{fig:neel}
\end{figure}
It is often convenient to express a spin-configuration not in terms of the spin directly, but in terms of the N\'eel-vector (see FIG. \ref{fig:neel}). In particular, in the experimentally observed \emph{antiphase} stripes, the N\'eel-vector changes sign between two adjacent charge stripes.
\item On a tetragonal lattice, in addition to the above, a \emph{checkerboard} pattern which respects the lattice rotation symmety is possible. Here the charge and spin-densities have equal modulations in both the $\hat x$ and $\hat y$ directions, i.e.
\begin{align}
 \langle \hat \rho(\vect r)\rangle = \rho_0 + \(\rho_{\vect Q_{c}}e^{i \vect Q_{c,1} \cdot \vect r}+\rho_{\vect Q_{c}}e^{i \vect Q_{c,2} \cdot \vect r}+ c.c.\).
\end{align}
and likewise for the spin-density. All of these ordering patterns are of relevance in the cuprates, and we will discuss them in turn.
\end{enumerate}

\section{Quantum melting of stripes: possible phase diagrams}\label{sec:phasedia}
Before addressing the quantum phase transition in detail, we will now briefly discuss a possible phase diagram and some properties of the phases which we consider. The limiting cases are, on the one hand, the usual Fermi-Liquid with a large Fermi surface which respects all symmetries of the underlying crystalline lattice. On the other hand, there are various striped phases where translation symmetry is broken by static spin and charge order as discussed in the previous section. In these phases the original large Fermi-surface is reconstructed. As a striped phase undergoes a transition (or a sequence of transitions) into the Fermi-Liquid, the symmetries broken by stripe-order are restored. Depending on the stripe order in question, these are spin-rotation, lattice rotation and lattice translations symmetry. In general, there may be intermediate phases where only a subset of the symmetries are restored. For example, when spin-rotation symmetry is restored but lattice symmetries remain broken, a spin-striped phase turns into a phase with charge-stripes only. Intermediate phases without stripe-order are also possible, e.g. a spin-nematic phase where lattice symmetries are restored but spin rotation symmetry remains broken.

\begin{figure}[ht]
\includegraphics[width=.7\columnwidth]{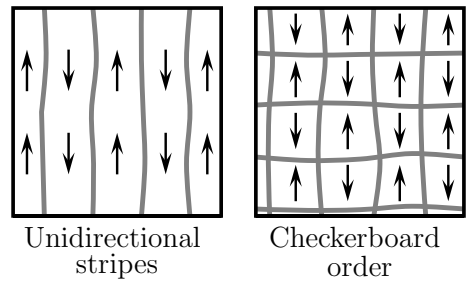}
\caption{Unidirectional and Checkerboard stripe patterns. The gray lines denote minima in the charge density and the arrows denote the orientation of the N\'eel vector.}
\label{fig:static}
\end{figure}

\subsection{Structure of defects in the stripe order parameter}
Just like melting of an ordinary crystal, the melting of stripes is most conveniently described in terms of defects in the stripe order. In the limiting case of perfect long-range stripe order, no defects are present. The Fermi liquid lies in the opposite limit, where stripe order is completely destroyed and the number of defects is no longer well defined, i.e. it may be viewed as a condensate of all possible defects. Intermediate phases can be realized (and characterized) by proliferating only a subset of allowed defects.

\begin{figure}[ht]
\includegraphics[width=.7\columnwidth]{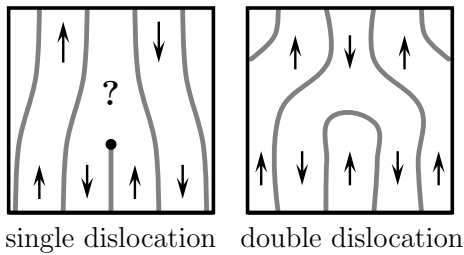}
\caption{Left: Single dislocations in the charge stripes are bound to half-dislocations for the spin-order, leading to frustration. Right: Double dislocations in the charge stripes avoid frustration. The gray lines denote minima in the charge density and the arrows denote the orientation of the N\'eel vector.}
\label{fig:dis}
\end{figure}

\begin{figure}
\includegraphics[width=\columnwidth]{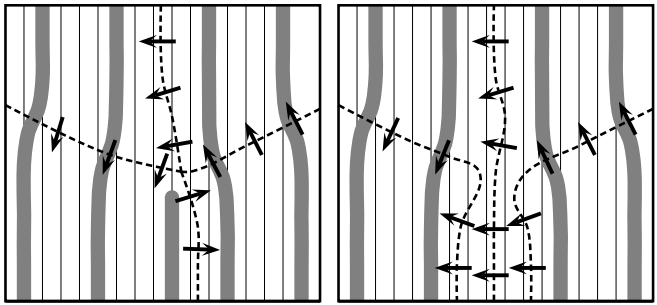}
\caption{Left: Four elementary directed domain walls (dashed lines) meet at a stripe dislocation. The charge order in neighboring domains is related by a primitive lattice translation. The thin vertical lines are drawn as a guide for the eye, and are separated by one lattice spacing. Right: Configuration with three domain walls but without a dislocation.}
\label{fig:walls}
\end{figure}

Before addressing the phase transition, it is necessary to identify the possible topological defects. We begin by considering incommensurate, unidirectional spin and charge stripes. To allow for defects we must allow the order parameters to vary in space and time. It is convenient to introduce the phases of the stripe-order parameters as
\begin{align}
\theta_{s,c}(\vect r,t)\equiv \[\vect Q_{s,c}(\vect r,t) - \vect Q _{s,c}(0,0)\] \cdot \vect r.
\end{align}
$\theta(\vect r, t)$ has the interpretation of being ($Q$ times) the local displacement of the stripes in the $\hat{Q}$-direction at time $t$. The fundamental topological defects are a) dislocations where the phase of the spin-order winds by $\pi$ and $\vec N$ rotates by $\pi$ and b) dislocations where the phase of the spin order $\theta_s$ winds by $2\pi$ (see FIG. \ref{fig:dis}). Around these defects the phase of the charge stripe $\theta_c=2\theta_s$ winds by $2\pi$ and $4\pi$, respectively. We will hence refer to defects of type a) as single dislocations and to defects of type b) as double dislocations. Spin order is frustrated around single dislocations but not around double dislocations (see FIG. \ref{fig:dis}). In the commensurate case of a period $m$ charge-stripe $\vect Q_c = \left(\frac{2\pi}{m}, 0 \right)$ we can also have elementary oriented domain walls where the entire stripe-pattern is shifted by one lattice spacing ($\theta_c$ increases by $\frac{2\pi}{m}$). A single dislocation is then located at the point where $m$ such domain walls meet (see Fig. \ref{fig:walls}).

In a conventional (weak coupling) approach, the stripe melting is described as an $XY$-transition for the charge-stripe order parameter $e^{i \theta_c}$. At this transition, \emph{all} dislocations proliferate and the resulting state is the Fermi liquid. However, in the presence of a Fermi surface, this approach becomes problematic for the reasons discussed in Section \ref{sec:problems}. It is then natural to ask whether these complications are still present at a modified transition where only certain dislocations proliferate.

A physical mechanism that may lead to a situation where some, but not all defects proliferate was already mentioned above. We have seen that when spin order is present, it is frustrated around single dislocations. This raises the energy cost of such dislocations\cite{footnote0}, potentially even above the energy of double dislocations which are not frustrated. Thermal melting by proliferating double dislocations was studied for spin stripes in Ref. \onlinecite{krueger} and for a striped superconductor/the Larkin-Ovchinnikov (LO) state in Refs. \onlinecite{bfk,leo}. Quantum melting of the LO state was discussed in a cold-atoms context in Ref. \onlinecite{leo2}.

 Even in the absence of long-range spin order, strong short-range spin correlations may significantly raise the core-energy of single dislocations. In this paper we thus explore a stripe melting transition where double dislocations proliferate, while single dislocations remain gapped. In the resulting stripe liquid phase, all lattice symmetries are restored. This phase was first proposed by Zaanen\cite{zaa} and co-workers and further explored in Refs. \onlinecite{demler,subir,zaanen}. Despite having the same symmetries, it is \emph{not} the regular Fermi liquid due to the gapped single dislocations. A clear distinction between the non-trivial stripe liquid phase and the Fermi liquid lies in the topological structure, i.e. the former has a ground state degeneracy on non-trivial manifolds. While this difference is completely sharp theoretically, it is extremely difficult to probe experimentally. In particular, the stripe liquid phase has the same large Fermi surface with well defined Landau quasiparticles as the weakly interacting Fermi liquid. Its single-electron properties, transport and low-energy thermodynamics are also identical, thus this phase may have easily been mistaken for a Fermi liquid. A detailed discussion of this phase along with a more physical picture is given in Section \ref{sec:slm}.

\begin{figure}
\includegraphics[width=.7\columnwidth]{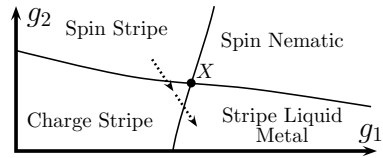}
\caption{Schematic zero-temperature phase diagram close to the multicritical point $X$ which separates the four phases discussed in the text. The dashed line is parametrized by $g$.}
\label{fig:sn2}
\end{figure}

Above we saw that the order parameters $b=e^{i \theta_s}$ and $\vec N$ are intertwined only around single dislocations, while around double dislocations they are independent. Thus at energies well below the core energy of single dislocations, $b$ and $\vec N$ become separately well defined (this will be explained more formally in Section \ref{sec:gauge}). We can envisage four different phases (see FIG. \ref{fig:sn2}), in all of which the single dislocations are gapped. First, $\langle \vec N \rangle \neq 0$, $\langle b\rangle \neq 0$ is the ordered phase with both spin and charge order. Second, $\langle \vec N \rangle = 0$, $\langle b\rangle \neq 0$ describes a phase with a charge stripes but without spin order. Third when $\langle \vec N \rangle \neq 0$, $\langle b\rangle = 0$, lattice symmetries and time reversal symmetries are restored. Thus in this phase the spin expectation value $\vec S =0$ but spin rotation symmetry is broken by a spin quadrupole moment $Q_{ab} = N_a N_b - \frac{1}{3}\vec N^2 \delta_{ab}$ (such a phase is frequently referred to as spin nematic). Finally $\langle b \rangle = 0, \langle \vec N \rangle = 0$ describes the non-trivial stripe liquid we mentioned above and which we discuss in detail in the following section.

 In the presence of itinerant fermions the two latter phases have a conventional large Fermi surface while in the striped phases with $\langle b \rangle \neq 0$ the Fermi surface is reconstructed by the stripe order.

\section{Stripe loop metals}\label{sec:slm}

\subsection{Physical picture}\label{sec:phys}
As we mentioned above, the distinction between the Stripe Liquid Metal obtained after proliferation of paired dislocations and the regular Fermi-Liquid is topological and thus difficult to detect. On a pictorial level, however, the difference between these phases becomes very intuitive, by looking at the charge density $\sim e^{i \vect Q_c \cdot \vect r}b^2$. In a perfectly stripe-ordered phase all the stripes extend across the entire system, and there are no stripe endings. In the regular Fermi liquid, all kinds of fluctuations are allowed, in particular there are dislocations where stripes end (see FIG. \ref{fig:slm}). We are instead interested in a stripe liquid phases where only double dislocations are allowed. Introducing any number of (static) paired dislocations into a stripe ordered phase turns the stripe pattern into a set of \emph{closed} stripe-loops. In the stripe liquid phase, where these dislocations proliferate, there will thus be a fluctuating pattern of stripe-loops, but no stripe endings (see FIG. \ref{fig:slm}).

\begin{figure}[ht]
\includegraphics[width=\columnwidth]{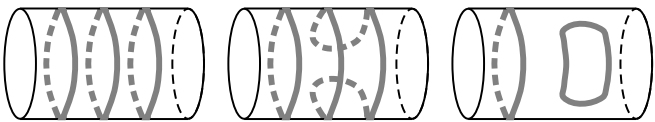}
\caption{Three smoothly connected configuration of the Stripe Loop Metal on a cylinder. The number of stripes winding around the cylinder is conserved \emph{mod 2}. The parity of closed stripes is a topological quantum number distinguishing between two possible ground states.}
\label{fig:top}
\end{figure}

Indeed, this picture can be made precise by connecting the stripe patterns to features of the field theory, which we will demonstrate below. To see the topological nature of the SLM it is useful to consider the system on a cylinder, around which the stripes wind in the ordered phase. In the SLM phase, all fluctuations of stripes are allowed, as long as there are no loose ends. Thus the number of \emph{closed} charge stripes winding around the cylinder is conserved \emph{mod 2} (see FIG. \ref{fig:top}). There are two ground states which are distinct by the parity $P_c$ of closed stripes around the cylinder.

 The distinction between these two states is clearly topological, i.e. cannot be determined by just looking at any finite region of the cylinder.

\subsection{Gauge theory}\label{sec:gauge}
To make the Stripe Loop representation of these phases more precise we now connect it to the corresponding field-theory. Formally to express the physical order parameters in terms of $b$ and $\vec N$, i.e.
\begin{align}
&\vec M =\vec N b\\
&\psi = b^2\label{eqn:fraccharge}
\end{align}
the phase $\theta_s$ of $b$ should fall in the interval $[0,\pi]$. It is more convenient to instead let $\theta_s \in[0,2\pi]$, which comes at the cost of a local $Z_2$ redundancy. At each site we can independently change the sign of both $b$ and $\vec N$ without affecting the physical order parameters. As this \emph{emergent} gauge degree of freedom is an artifact of our chosen representation of the operators, any physical operator must be a gauge-invariant combination of these operators. In particular, inverting Eq. (\ref{eqn:fraccharge}) leads to
\begin{align}
 b_{\vect r}=s_{\vect r}\sqrt{\psi_{\vect r}} = s_{\vect r}e^{i \theta_c(\vect r)/2}\label{eqn:bgauge},
\end{align}
where $s_{\vect{r}}$ depends on a particular gauge, but the combination $b_{\vect r}s_{\vect r}$ is gauge invariant, i.e. under a gauge transformation where $b_{\vect r}\rightarrow -b_{\vect r}$ we also have $s_{\vect{r}} \rightarrow - s_{\vect{r}}$.

\begin{figure}[ht]
\includegraphics[width=\columnwidth]{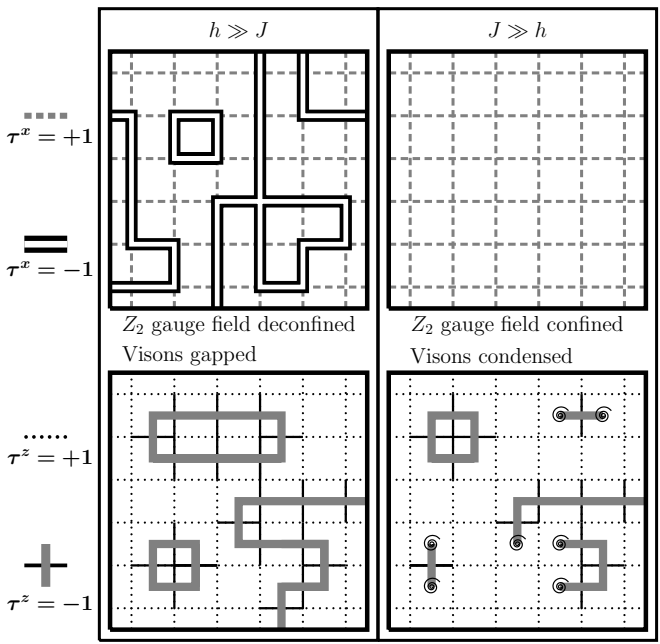}
\caption{Left: Typical configuration in the \emph{deconfined} phase of $Z_2$ gauge theory. Electric field lines, $\tau^x_{ij}=-1$, denoted by double black lines in the upper picture, are deconfined. Visons - plaquettes with an odd number of $\tau^z_{ij}=-1$ - denoted by a spiral in the lower picture, are absent. Visons act as sources for the thick gray lines, thus there are only closed strings of these.
Right:
 Typical configuration in the \emph{confined} phase. The electric field is confined and visons have proliferated, thus there are both open and closed gray lines. In the present context the gray lines describe charge stripes.}
\label{fig:vis}
\end{figure}

 Any term in the Hamiltonian of $b$ and $\vec N$ needs to be invariant under such a transformation, in particular the kinetic terms must take the form
\begin{align}
 H_\text{hopping} = -t_b\sum_{\langle\vect r\vect r'\rangle} b^*_{\vect r} \tau^z_{\vect r,\vect r'}b_{\vect r'}-t_{\vec N}\sum_{\langle\vect r\vect r'\rangle} \vec N_{\vect r} \tau^z_{\vect r,\vect r'}\vec N_{\vect r'}.
\end{align}
Here $\tau^z_{\vect r,\vect r'}$ is a $Z_2$ gauge field, which transforms under a local $Z_2$ gauge transformation $b_{\vect r}\rightarrow \sigma_{\vect r}b_{\vect r}$, $\vec N_{\vect r}\rightarrow \sigma_{\vect r}\vec N_{\vect r}$ as $\tau_{\vect r,\vect r'}\rightarrow \sigma_{\vect r}\sigma_{\vect r'}\tau_{\vect r,\vect r'}$, with $\sigma_{\vect r} \in \{+1,-1\}$. While $b$ itself is not gauge-invariant, the integral of its phase $\theta_s$ around a close loop is, i.e.
\begin{align}
&e^{i 2\pi n_s}=\(\tau^z_{\vect r,\vect r_1}\tau^z_{\vect r_2,\vect r_3}\ldots \tau^z_{\vect r_N,\vect r}\)e^{i\pi n_c}\label{eqn:loops}\\
&\oint \vect \nabla \theta_{s,c} \cdot d\vect s=2\pi n_{s,c}.
\end{align}
In the ordered phase $\langle b \rangle \neq 0$, single valuedness of $b$ requires that $n_s$ be an integer for any loop. We now consider a loop enclosing a single defect where $\theta_c$ winds by $2\pi$, i.e. $n_c=1$. Then Eq. (\ref{eqn:loops}) requires that around this loop
\begin{align}
\tau_{\vect r,\vect r_1}\tau_{\vect r_2,\vect r_3}\ldots \tau_{\vect r_N,\vect r}=-1.
\end{align}
We have seen that single dislocations in $e^{i\theta_c}$ corresponds to a $Z_2$ gauge flux of $\pi$ in the gauge theory. Such an excitation (see FIG. \ref{fig:vis}) is usually referred to as a vison. Therefore the transition between Fermi Liquid and the SLM is described as a confinement transition of the $Z_2$ gauge theory, where visons are gapped in the \emph{deconfined} phase (SLM) while they proliferate in the \emph{confined} phase (FL). At the deconfinement transition, both $b$ and $\vec N$ are massive and do not significantly affect the critical properties of the gauge theory. However, the phase of the gauge theory has important consequences for the $b$ and $\vec N$ fields. As long as the visons are gapped, $b$ and $\vec N$ are independently well defined. Once the visons condense $b$ and $\vec N$ become confined, i.e. they are glued together into $Z_2$-charge neutral objects like $b^2$ and $b\vec N$.

 Let us briefly review some basic properties of $Z_2$ gauge theory: A simple Hamiltonian for the $Z_2$ Gauge theory on a square lattice is

\begin{align}
 H=- J \sum_{\langle ij \rangle} \tau^x_{i,j} - h \sum_\Box\prod_{\langle ij \rangle \in\Box} \tau^z_{i,j},
\end{align}
where $\Box$ denotes a square plaquette. In addition, the system is subject to the gauge constraint
\begin{align}
\hat G_i = \tau^x_{i,i+\hat x}\tau^x_{i,i-\hat x}\tau^x_{i,i+\hat y}\tau^x_{i,i-\hat y} =1.
\end{align}
For $h \rightarrow 0$ (the \emph{confined} phase of the gauge theory) the ground state is
\begin{align}
 |0\rangle &= \prod_{\langle ij\rangle} |\tau_{ij}^x =+1\rangle
&\sim\prod_{\langle ij\rangle} \(|\tau_{ij}^z =+1\rangle+|\tau_{ij}^z =-1\rangle\).
\end{align}
In the \emph{deconfined} phase, $J \rightarrow 0$, it is given by
\begin{align}
|0\rangle=\(1+ \sum_i G_i+ \sum_{i\neq j} G_{i}G_{j}+\ldots\)\prod_{\langle ij\rangle} |\tau_{ij}^z=+1\rangle.
\end{align}
Pictorially this state can be represented by coloring in gray each bond of the \emph{dual} lattice which crosses a link where $\tau_{ij}^z =-1$ (see Fig.\ref{fig:vis}). The confined ground state is then a superposition of all possible configuration with both open and closed gray lines at low energies.
The vison number is given by the gauge flux $n=\prod_{\Box}\tau^z$ thus there is a vison at the end of each open line. In the deconfined phase (of the $Z_2$ gauge field) the visons are gapped, thus there are only closed gray lines. As visons (i.e. charge stripe endings) act as sources for these gray lines, these lines can be identified with the charge stripes.\\
\subsection{SLM on tetragonal lattices}\label{sec:tetslm}
\begin{figure}[ht]
\includegraphics[width=.8\columnwidth]{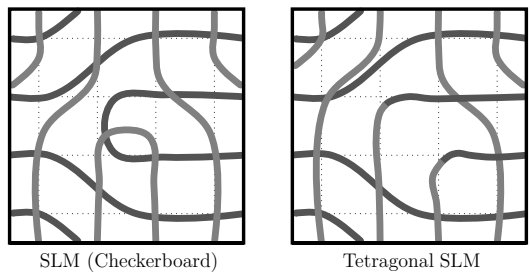}
\caption{Left: Checkerboard pattern with a double dislocation in both the vertical and the horizontal stripe. Right: There is a different kinds of dislocation, where a horizontal stripe turns into a vertical one. Such dislocations also do not suffer from spin-frustration, thus there is the possibility of a phase where both the double dislocations and this new kind of dislocations proliferate.}
\label{fig:tet}
\end{figure}

The previous dicussions considered unidirectional charge order. On tetragonal lattices there is the additional possibility that the order respects the lattice rotation symmetry, i.e.
\begin{align}
 \langle \hat \rho(\vect r)\rangle = \rho_0 + \(\rho_{\vect Q_{c}}e^{i Q_{c,x} x}+\rho_{\vect Q_{c}}e^{i Q_{c,y} y}+ c.c.\).
\end{align}
Now there are two possibilities of a Stripe Loop Metal respecting lattice rotation symmetries. The SLM can be realized by independent fluctuations of the vertical and horizontal stripes, i.e. we consider double dislocations in $\theta_{c,x}$ and $\theta_{c,y}$ separately. Such a state is depicted in FIG. \ref{fig:tet} (left). In such a phase there are two kinds of (gapped) visons, corresponding to endings of horizontal or vertical stripes.

There is a further possibility for a SLM phase, distinct from the previous case, which is illustrated in FIG. \ref{fig:tet} (right). This corresponds to allowing, in addition to the double dislocations in $\theta_{c,x}$ and $\theta_{c,y}$ separately, a new kind of defect where both the phases of \emph{both} $\theta_{c,x}$ and $\theta_{c,y}$ wind by $\pm 2\pi$. It is easy to see that in the presence of spin order, such dislocations are not frustrated.

In order to describe the transition we begin with two equal order parameters $\langle \psi_{x}\rangle=\langle \psi_{y}\rangle\neq 0$. To include the new kind of defect, we write the stripe order parameter in a somewhat modified form
\begin{align}
&\psi_x = b_1 b_2\nonumber \\
&\psi_y = b_1 b_2^*,\label{eqn:tetrafrac}.
\end{align}
 Now at a defect where the phase of $b_1$ winds by $2\pi$, \emph{both} $\theta_{c,x}$ and $\theta_{c,y}$ wind by $2\pi$. Similarly, around a single dislocation in $b_2$, $\theta_{c,x}$ winds by $2\pi$ while $\theta_{c,y}$ winds by $-2\pi$. Further, at a single dislocation in both $b_1$ and $b_2$, the phase of $\psi_x$ winds by $4\pi$ while the phase of $\psi_y$ does not wind at all. Thus this decomposition captures both the double dislocations discussed above, as well as the new kind of dislocation. Formally this decomposition has only a single $Z_2$ redundancy, associated with changing the sign of both $b_1$ and $b_2$ together.

\section{Charge stripe melting transitions}\label{sec:chargemelting}
\subsection{Orthorhombic crystal}
We begin with the simplest case, i.e. unidirectional charge order without spin-order. Our strategy will be to initially consider incommensurate stripes (i.e. no pinning of the stripe order parameter to the crystalline lattice) in the absence of an electronic Fermi surface. We will justify this \emph{a posteriori} by showing that both these couplings are irrelevant (in the RG sense). Both in the ordered phase and in the SLM the $Z_2$ fluxes (equivalently single dislocations) are gapped, so we can neglect them when studying this transition. To determine the universal properties of the phase transition it is convenient to employ a `soft-spin' formulation. The low-energy effective field theory is given by the most relevant, symmetry-allowed terms of $b, \vec N$, i.e.
\begin{align}
S[b, \vec N] & = \int d\tau d^2r {\cal L}_b + {\cal L}_N + {\cal L}_{bN}\label{eqn:ortho} \\
{\cal L}_b & = |\nabla b|^2 + \frac{1}{v_c^2}|\partial_\tau b|^2 + r_b |b|^2 + u_b |b|^4 \\
{\cal L}_N & = |\nabla \vec N|^2 + \frac{1}{v_s^2}|\partial_\tau \vec N|^2 + r_N |\vec N|^2 + u_N|\vec N|^4 \\
{\cal L}_{bN} & = v|b|^2 |\vec N|^2.
\end{align}
We note in particular that under a combined time reversal and spatial inversion (along $\vect Q$), $b$ is invariant, thus no linear derivative terms are allowed. When the spin is disordered, $\vec N$ is gapped and may be integrated out. What remains is then a 3D $XY$ transition for the $b$ field. The critical properties of this model are well known\cite{nuxy}. It should be noted that the transition can be formally described as a regular XY transition for the $b$ field, although the physical stripe melting transition is in a fundamentally different universality class. The fundamental field $b$ itself is not a physical field, but rather the stripe order parameter $\psi=b^2$. More generally, only operators invariant under the local $Z_2$ transformations are physical, thus the operator content of this universality class is very different from the usual 3D $XY$ transition. However, the critical properties are easily derivable from the ones of the 3D $XY$ universality class. The anomalous dimension of the composite field\cite{sudupe} $\psi$, i.e. $\eta_\psi \approx 1.49$ is dramatically different from the anomalous dimension of $b$, $\eta_b \approx 0.04$. This universality class has been studied in Refs. \onlinecite{tsfisher,tarun,isakov} where it was dubbed $XY^*$.

Let us now consider the stability of the $XY^*$ stripe melting transition to various perturbations.
\begin{enumerate}
\item
{\em Pinning to lattice}

The pinning of the period-4 charge stripe to the lattice is described by a term
\begin{equation}
S_\text{pin} = -\lambda \int d\tau d^2r \cos(4\theta_c)
\end{equation}
In terms of the phase $\theta_s$ of the $b$-field this becomes
\begin{equation}
S_\text{pin} = - \lambda \int d\tau d^2r \cos(8 \theta_s)
\end{equation}
This is an $8$-fold anisotropy on the XY field at the $2+1$-D XY fixed point which is well known to be strongly irrelevant. Thus despite the commensurate ordering wavevector the stripe fluctuations de-pin from the lattice as the quantum critical point is approached. Clearly this pinning term is important at long distances in the stripe ordered state, and thus represents a dangerously irrelevant perturbation of the critical fixed point. If the stripe ordering occurs at incommensurate wavevector then there is no pinning of the stripes either at the critical point or in the ordered phase.

\item
{\em Coupling to Fermi surface}

Thus far we have ignored the presence of the conduction electron Fermi surface. It couples to the critical stripe fluctuations in several ways which we now analyze. The most interesting coupling arises in the case where the stripe ordering wavevector connects two points of the Fermi surface. In the stripe ordered phase this leads to a term
\begin{equation}
\label{stripecouple}
\kappa \psi \sum_k c^\dagger_{k+ Q} c_{k} + h.c.
\end{equation}
in the conduction electron Hamiltonian which will reconstruct the electronic Fermi surface.
At the critical point or in the stripe melted phase this coupling will lead to a damping of the stripe fluctuations. Integrating out the conduction electrons in the stripe melted phase leads to the standard Landau-damping term
\begin{equation}
\label{ldcs}
\lambda_d \int d\omega d^2 q |\omega| |\psi (q,\omega)|^2
\end{equation}
This may be viewed as an interaction (between the $b$ fields) that is long ranged in imaginary time between the stripe fluctuations. The relevance/irrelevance of this term at the stripe melting $XY^*$ critical point is readily ascertained by power-counting. Under a renormalization group transformation $x \rightarrow x' = \frac{x}{s}, \tau \rightarrow \tau' = \frac{\tau}{s}$, we have
$\psi \rightarrow \psi' = s^{\Delta} \psi$ with $\Delta = \frac{1+ \eta}{2}$. This implies
\begin{equation}
\lambda_d' = \lambda_d s^{1 - \eta}
\end{equation}
As $\eta > 1$ at the $XY^*$ fixed point, the Landau damping of the critical stripe fluctuations is irrelevant.

A direct coupling of the energy density of the stripe fluctuations to soft shape fluctuations of the Fermi-surface (in the continuum only to the breathing mode) is also allowed by symmetry. However, it was shown in Ref. \onlinecite{subir} that if the correlation length exponent $\nu_{XY} > \frac{2}{3}$, such couplings are irrelevant. For the 3D $XY$ model\cite{nuxy}, $\nu=0.67155$, so this coupling is also irrelevant.

For an orthorhombic crystal all other couplings to the Fermi surface are even more strongly irrelevant. The situation with tetragonal symmetry is more complicated and will be analyzed below.

\item
{\em Fermi surface reconstruction}

In the stripe ordered phase gaps will open up at `hot spots' of the original large Fermi surface which are connected to each other by the ordering wavevector $\vect Q$. The irrelevance of the coupling of the critical stripe fluctuations to the conduction electrons suggests that the scale at which the Fermi surface reconstructs is parametrically lower than the scale at which the stripe order onsets. To explore this we now describe the coupled system of stripes and conduction electrons in a different framework that does not integrate out the conduction electrons. We restrict attention to conduction electron states near the hot spots and linearize their dispersion. The resulting action takes the form
\begin{align}
&S = \int d\tau d^2r {\cal L}_c + {\cal L}_{\text{int}} + {\cal L}_b \label{eqm:fermionstripe}\\
&{\cal L}_c = \sum_i \bar{c}_i\(\partial_\tau -i\vect v_i \cdot \vect \nabla\)c_i \nonumber \\
&{\cal L}_{\text{int}} = \kappa\sum_{ij}\left(\psi_{ij} \bar{c}_i c_j + c.c. \right) \nonumber,
\end{align}
where $c_i$ denotes a fermion at the $i$th hotspot, $\vect v_i$ is the Fermi velocity at the $i$th hotspot and $\psi_{ij}$ is a stripe-order parameter with a wave-vector that connects the $i$th and the $j$th hotspot. Here we generalized the fermion action to allow for more than one stripe-ordering wavevector, as will be the case in the presence of tetragonal symmetry, which we will discuss below.
 ${\cal L}_b$ is the Euclidean Lagrangian of the $b$ field(s) in the $2+1$-D XY universality class.

Consider now the RG transformation appropriate for the $3D$ XY fixed point. The action $S_c = \int d\tau d^2r {\cal L}_c$ is invariant under this RG transformation provided that we scale
\begin{equation}
c_{1,2} \rightarrow c'_{1,2} = s c_{1,2}
\end{equation}
With this scaling the full action is at a `decoupled' fixed point at $ \kappa = 0$. By power counting we see that a small $\kappa$ renormalizes as
\begin{equation}
\kappa' = \kappa s^{\frac{1-\eta}{2}}\label{eqn:aflow}
\end{equation}
As $\eta > 1$ $\kappa$ is irrelevant consistent with the analysis of the previous section. Now assume we are in the ordered phase a distance $\delta$ from the critical point. The energy scale $\Delta_\text{stripe}$ of stripe formation ({\em i.e} the scale at which the critical theory ${\cal L}_b$ crosses over from the critical to the ordered fixed point) increases as
\begin{equation}
\Delta_\text{stripe} \sim |\delta|^{\nu z}
\end{equation}

\begin{figure}[ht]
 \includegraphics[width=6.5cm]{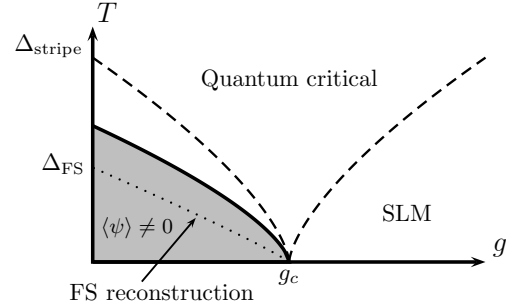}
\caption{Stripe ordering and reconstruction of the Fermi-surface occur at parametrically different energy scales. The dashed lines $\Delta_\text{stripe} \sim |g-g_c|^{0.67}$ bound the quantum critical region and mark the onset of stripe order for $g <g_c$. The solid line representing the finite-temperature phase transition is parametrically the same as $\Delta_\text{stripe}$. The dotted line denotes the scale where the Fermi-surface reconstructs and is given by $\Delta_\text{FS} \sim |g-g_c|^{0.83}$.}
\end{figure}

To determine the value of the gap at the hot spots, we consider the self-energy of the electrons. It must have the dimensions of energy, i.e.
\begin{align}
\Sigma\(\omega , \kappa, \Delta_{\text{stripe}}\)= \frac{1}{s}\Sigma \(s\omega,s^{(\eta-1)/2} \kappa ,s \Delta_\text{stripe}\).
\end{align}
We choose $s$ such that $s\Delta_\text{stripe}=1$ and scale by a factor of $\kappa^2\Delta_\text{Stripe}^{\eta}/\omega $, obtaining
\begin{align}
\Sigma\(\omega , \kappa, \Delta_{\text{stripe}}\)=\frac{\(\kappa \Delta_\text{Stripe}^{(1+\eta)/2}\)^2}{\omega} f\(\frac{\omega}{\Delta_\text{stripe}}, \frac{\kappa}{\Delta_\text{stripe}^{(1-\eta)/2}}\)\label{eqn:scaleselfenergy},
\end{align}
where $f$ is a universal scaling function. On the ordered side, close to the hot-spot, the electron self-energy has the form (see FIG. \ref{fig:recdia}).
\begin{align}
\Sigma(\omega ,k) = \frac{\Delta_\text{FS}^2}{i\omega - \epsilon_{\vect k + \vect Q}},
\end{align}
where $\Delta_\text{FS}$ is the size of the gap at the hot spots.
\begin{figure}[ht]
\includegraphics[width=6cm]{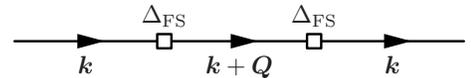}
\caption{Diagram for the fermion self-energy in the ordered phase.}
\label{fig:recdia}
\end{figure}

By matching this to Eq. (\ref{eqn:scaleselfenergy}) we get
\begin{align}
\Delta_\text{FS}\sim \Delta_\text{Stripe}^{(1+\eta)/2}.
\end{align}
Thus as expected the hot spot gap $\Delta_{FS}$ (the scale at which the Fermi surface reconstructs) is parametrically different from the scale at which stripe ordering occurs. The two energy scales have a ratio
\begin{equation}
\frac{\Delta_{FS}}{\Delta_\text{stripe}} \sim\Delta_\text{Stripe}^{ (\eta-1)/2}
\end{equation}
which goes to zero as the critical point is approached.
\begin{figure*}
 \includegraphics[width=14cm]{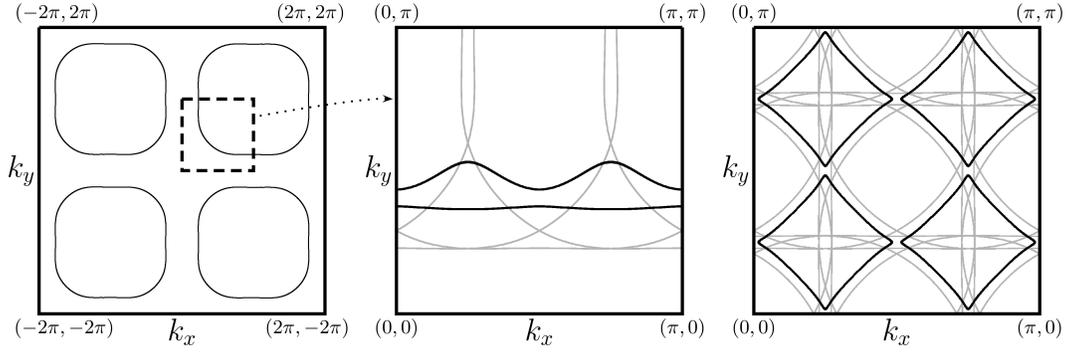}
\caption{Reconstruction of the Fermi-surface due to period 4 charge stripes. On the left, the unreconstructed large Fermi surface is shown in the extended zone scheme. The dashed lines frame the part of the Brillouin zone that is used to show the reconstructed Fermi-surface in the center and on the right. In the center reconstructed Fermi-surface due to uni-directional charge stripes is shown in black, while the gray lines indicate the original, folded Fermi surface. On the right, the reconstructed Fermi-surface due to checkerboard order is displayed in similar fashion.}\label{fig.reconstruction}
\end{figure*}

\end{enumerate}
Thus the $XY^*$ stripe melting critical point survives unmodified by either the coupling to the lattice or to the electronic Fermi surface. It therefore provides a concrete tractable example of a stripe melting transition in a metal. On tuning through this transition the Fermi surface undergoes a reconstruction (see Fig. \ref{fig.reconstruction}. For a more detailed discussion including different parameters see Ref. \onlinecite{sebastian}). This of course happens through the coupling of the stripe order parameter to the conduction electrons as described above in Eqn. \ref{stripecouple}.

\subsection{Tetragonal crystal}
\begin{figure}
\includegraphics[width=8cm]{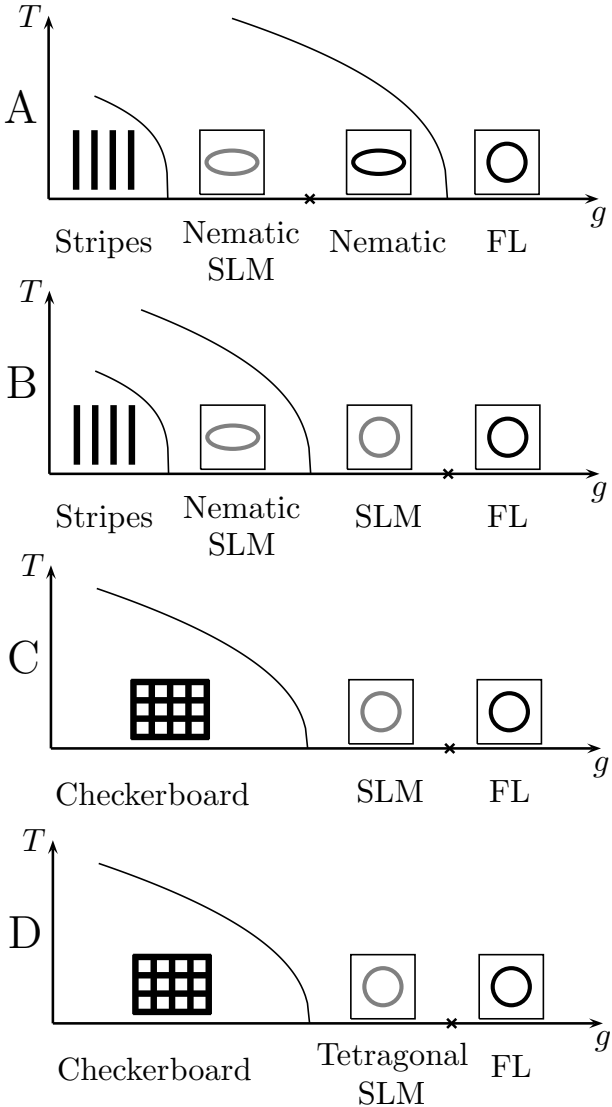}
\caption{Finite temperature phase diagrams for the three sequences of phase transitions which we discuss here.}
\label{fig:seq2}
\end{figure}

We now turn to the situation where in addition to lattice translations, we also have fourfold rotation symmetry.
In this case, the average charge density $\rho$ will be modulated in both the $x$ and the $y$ direction, i.e.
\begin{equation}
\rho_r =\rho_0+ \sum_{i=x,y}e^{2i {\bf Q}_i\cdot {\bf r}} \psi_i + e^{-2 i {\bf Q}_i\cdot {\bf r}} \psi_i^*.
\end{equation}
%The order parameters $\psi_s$ transform under the lattice symmetries as
%\begin{align}
%&T_x : \psi_x \rightarrow \psi_x e^{\i 2 Q_x} , \ \ \ \psi_y \rightarrow \psi_y \\
%&T_y : \psi_x \rightarrow \psi_x , \ \ \ \psi_y \rightarrow \psi_y e^{\i 2 Q_y} \\
%&R_{\frac{\pi}{2}}: \psi_x \rightarrow \psi_y , \ \ \ \psi_y \rightarrow \psi_x^* \\
%&P_{xy}: \psi_y \rightarrow \psi_x , \ \ \ \psi_x \rightarrow \psi_y
%\end{align}
%where $T_s$ are translations, $R_{\frac{\pi}{2}}$ is a rotation by $\pi/2$ and $P_{xy}$ is the reflection about the diagonal. Further, under time reversal $\mathcal T$ we have
%\begin{align}
%\mathcal T: \psi_s \rightarrow \psi_s^*.
%\end{align}
For the translation-symmetry broken state there are now two possibilities: Rotational symmetry can be spontaneously broken $\langle \psi_x\rangle\neq 0$, $ \langle \psi_y\rangle=0$ (or vice versa), resulting in uni-directional stripes. Alternatively, the charge order can preserve rotational symmetry $\langle \psi_x\rangle= \langle \psi_y\rangle\neq 0$, resulting in a ``checkerboard'' pattern. As we will show below, the checkerboard state admits a direct transition into the Stripe Loop Metal. In order to reach the Fermi-Liquid, both translational and rotational symmetry need to be restored. We consider the four possible sequences of phase transitions which are shown in Fig. \ref{fig:seq2}.
\begin{enumerate}[label=\emph{\Alph*}. ]
 \item \emph{ Stripe $\rightarrow$ Nematic Stripe Loop Metal $\rightarrow$ Nematic $\rightarrow$ Fermi Liquid}\label{sequencea}\\
First we consider a phase transition between a unidirectional stripe-ordered state and a nematic Fermi-liquid. In this case one of the $\psi_i$ is zero on both sides of the transition. Furthermore, rotational symmetry remains broken after translation symmetry is restored, so we can describe the translation-symmetric state as a condensate of pairs of unidirectional stripe dislocations. We thus fractionalize the non-zero order parameter as in the orthorhombic case and obtain the same action as in Eq. \ref{eqn:ortho}. The properties of this phase transition are thus identical with the orthorhombic case which we discussed in the beginning. After translation symmetry is restored, we still have rotational order and the $Z_2$-structure remaining in a state which we dub Nematic SLM. The $Z_2$-structure disappears in a second order phase transition\cite{subir} to a nematic Fermi liquid, which is probably first order\cite{rahulmax}. Finally the theory of the phase transition between a nematic metal and a rotationally symmetric metal\cite{maxnematic} has a controlled limit\cite{nepsilon}, thus the properties of the entire sequence of phase transitions are understood.
 \item \emph{ Stripe $\rightarrow$ Nematic SLM$\rightarrow$ SLM $\rightarrow$ FL}\\
We again consider the same stripe-melting transition into the Nematic SLM, as discussed in the previous case. Instead of killing the topological structure, the system may first undergo a transition where rotational symmetry is restored. This transition is unaffected by the presence of the $Z_2$ structure and is again described by the nematic QCP discussed above. The last transition, where the topological structure is killed, is slightly different than in the case \ref{sequencea}, due to tetragonal symmetry. This scenario was also studied in Ref. \onlinecite{subir}, and the transition is expected to be first order.

 \item \emph{ Checkerboard$\rightarrow$ SLM $\rightarrow$ FL}\\
We now discuss the case where the ordered phase is described by two equal order parameters $\langle \psi_x\rangle=\langle \psi_y\rangle\neq 0$. To describe a transition out of this state we now fractionalize both order parameters
\begin{align}
 \psi_i = b_i^2.
\end{align}
We thus obtain a theory of two order parameters coupled to two fluctuating $Z_2$-gauge fields. In the ordered phase, both horizontal and vertical stripes coexist, and defects in the horizontal stripes and in the vertical stripes are decoupled. Again our aim is to melt the stripes by proliferating pairs of dislocations, without closing the gap for single dislocations. In that case we can safely ignore the gauge-field and describe the transition by the following Euclidean action:
\small
\begin{align}
S=& \sum_{i=x,y}\int_{\tau,\vect r} |\partial_\mu b_i|^2 + r |b_i|^2+u|b_i|^4+2v \int_{\tau,\vect r} |b_x|^2 |b_y|^2 \label{tetraaction}.
\end{align}
\normalsize
We first analyze this action in the absence of any coupling to the Fermi-surface and for $v=0$. This describes two decoupled $XY^*$ models which have a fixed point where $u$ is of order $\epsilon = 4-d$, where $d$ is the number of spacetime dimensions. The scaling dimension of $v$ close to this fixed point is given by
\begin{align}
\text{dim}[v]=d-\text{dim}[|b_x|^2]-\text{dim}[|b_y|^2]=\frac{2}{\nu_{O(2)}}-d\label{vscaling},
\end{align}
where $\nu_{O(2)}\approx 0.67155$ is the correlation length exponent of the $XY$-model in three dimensions\cite{nuxy}. We thus obtain
\begin{align}\text{dim}[v]\approx -0.02<0,\end{align} i.e. this perturbation is (weakly) irrelevant and the decoupled fixed point is stable. Next, tetragonal symmetry allows a linear coupling between the nematic fluctuations of the order parameter $O_\text{Nematic}(\vect r,\tau)=|b_x|^2-|b_y|^2$ and the ones of the Fermi surface
\begin{align}\rho_\text{nem.} \sim \sum_{\vect k} (\cos k_x - \cos k_y) c^\dagger_{\vect k} c_{\vect k},\end{align}
where $c^\dagger_k$ creates an electron with momentum $k$. Unlike the coupling in Eq.(\ref{stripecouple}), this term involves fermions close to a single point on the Fermi surface which leads to enhanced damping $|\omega|\rightarrow |\omega|/|q|$, i.e.
\begin{align}
\lambda_\text{Nematic}\int d^2q\int d\omega \frac{|\omega|}{|q|}\left|O_\text{Nematic}(q,\omega)\right|^2\label{nemterm}.
\end{align}
To determine its relevance, we note that this term couples different space-time points, thus its scaling dimension is determined by the dimension of $|b_x|^2-|b_y|^2$. But at the decoupled fixed point this is given by the dimension of $|b_x|^2$ alone, i.e.
\begin{align}
\text{dim}[\lambda_\text{Nematic}]=d-2\text{dim}[|b_x|^2-|b_y|^2]=\text{dim}[v].
\end{align}
Thus the coupling to the Fermi surface is also irrelevant in this case. Therefore the two $XY^*$ order parameters decouple both from each other and from the Fermi-surface at the phase transition and we obtain a second order transition from the checkerboard ordered phase into the SLM phase.
\item \emph{ Checkerboard$\rightarrow$ Tetragonal SLM $\rightarrow$ FL}\\
The final sequence we discuss here involves the tetragonal SLM introduced in Section \ref{sec:tetslm}. The order parameters of charge stripes in the $x$ and $y$ direction are decomposed as
\begin{align}
&\psi_x = b_1 b_2\nonumber \\
&\psi_y = b_1 b_2^* \label{eqn:tetdecomp},
\end{align}
which comes with a single $Z_2$ redundancy. To describe the transition, we thus disorder both $\psi_x$ and $\psi_y$ while keeping the gap of the single type of vison finite. The transformation properties of $b_1$, $b_2$ under lattice symmetries are shown in TABLE \ref{tab:sym}

\begin{table}[h]
 \caption{Transformation properties of various fields under discrete symmetries (we here assume period-4 charge stripes)}
\label{tab:sym}
 \begin{tabular}[c]{|l|c|c|c|c|}
\hline
Symmetry&$\psi_x$&$\psi_y$&$b_1$&$b_2$\\ \hline
$x$-Translation&$e^{i \pi/2}\psi_x$&$\psi_y$&$e^{i \pi/4}b_1$&$e^{i \pi/4}b_2$\\
$y$-Translation&$\psi_x$&$e^{i \pi/2}\psi_y$&$e^{i \pi/4}b_1$&$e^{-i \pi/4}b_2$\\
$\pi/2$ Rotation&$\psi_y$&$\psi_x^*$&$b_2^*$&$b_1$\\
Time Reversal &$\psi_x^*$&$\psi_y^*$&$b_1^*$&$b_2^*$\\
$x$-axis Reflection&$\psi_x$&$\psi_y^*$&$b_2$&$b_1$\\ \hline
 \end{tabular}
\end{table}

On symmetry grounds, this transition is described by Eq. (\ref{tetraaction}) and most of the above discussion follows through. In particular, the critical stripe fluctuations are governed by the same critical exponents, and the coupling to a Fermi surface is irrelevant. We note that in this representation $|\psi_x|^2-|\psi_y|^2=0$, thus it appears as if nematic fluctuations are strictly zero. However, the operator
\small
\begin{align}
&\sum_{\mu =x,y}\text{Im}\[b_1^*(\vect r+a\hat \mu) b_1(\vect r) \]\text{Im} \[b_2^*(\vect r+a\hat \mu) b_2(\vect r) \]\\
&=\frac{1}{2}\sum_{\mu =x,y} \text{Re}\[\psi^*_x(\vect r)\psi_x(\vect r+a\hat \mu)\]-\text{Re}\[\psi^*_y(\vect r+a\hat \mu)\psi_y(\vect r)\],
\end{align}
\normalsize
has the same symmetries as $|\psi_x|^2-|\psi_y|^2$. Thus the decomposition (\ref{eqn:tetdecomp}) does allow for nematic fluctuations.
 \end{enumerate}
\section{Spin stripe melting transitions}\label{sec:spin1}
For strong coupling stripe melting transitions it is natural to expect that the spin stripe order will melt through two phase transitions - first the spin order goes away while charge stripe order persists ( {\em i.e} translation symmetry remains broken) followed by a second transition where the charge stripe also melts. So far we focused exclusively on the second transition without addressing the first. It turns out that this transition (between a spin-stripe and a charge-stripe phase) is quite complicated and we currently do not have a good theory to describe it. To see the source of the difficulty, let us attempt to follow the same prescription as above, i.e. initially ignore the coupling to the Fermi-surface. Since $b$ is condensed, $\vec N$ is now physical (gauge-invariant), thus the transition is just the $O(3)$ transition with $\eta_N \approx 0.04$. Now if we add the conduction electrons, $\vec N$ can couple directly to their spin-density
\begin{equation}
\kappa_s \vec N\cdot \sum_k c^\dagger_{k+ Q}\vec\sigma c_{k} + h.c.
\end{equation}
Following the same argument as below Eq. (\ref{stripecouple}) we see that this is strongly relevant since $\eta_N \ll 1$ (the marginal case is $\eta=1$), and the fluctuating spin-stripes no longer dynamically decouple from the Fermi surface. The resulting theory was studied by Abanov and Chubukov \cite{abanov1,abanov2} and by Metlitski and Sachdev \cite{maxsdw}, who showed that it is strongly coupled at low energies. There is currently no controlled description of this transition.

\subsection{Stripe multicriticality}\label{sec:spin2}
\begin{figure}
\includegraphics[width=\columnwidth]{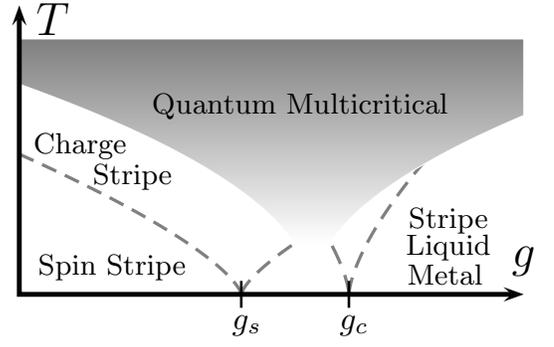}
\caption{Schematic finite temperature phase diagram as a function of $g$ (see FIG. \ref{fig:sn2}). The spin-order vanishes first at $g_s$ while the charge-order persists up to $g_c$.}
\label{fig:sn1}
\end{figure}

The difficulties with spin-stripe melting can be avoided if the spin-melting and the charge-melting transitions happen for nearby values of the tuning parameter $g$ (see FIG. \ref{fig:sn2}), which is likely in the cuprate phase diagram. Then, despite the presence of two distinct quantum phase transitions the somewhat higher-$T$ physics will be controlled by a ``mother" multicritical point where spin and charge stripe order simultaneously melt (see FIG. \ref{fig:sn1}). In our earlier work\cite{short}, we postulated that the temperature regime probed in the experiments of Ref. \onlinecite{AEPPLI} is controlled by such a multicritical stripe melting fixed point.

To study this multicritical point where both $b$ and $\vec N$ are critical, we again begin by ignoring the presence of the Fermi surface and the underlying lattice. The effective theory for this transition is the same as in (Eq. \ref{eqn:ortho}) which we reproduce here for convenience:
\begin{align}
S[b, \vec N] & = \int d\tau d^2r {\cal L}_b + {\cal L}_N + {\cal L}_{bN} \\
{\cal L}_b & = |\nabla b|^2 + \frac{1}{v_c^2}|\partial_\tau b|^2 + r_b |b|^2 + u_b |b|^4 \\
{\cal L}_N & = |\nabla \vec N|^2 + \frac{1}{v_s^2}|\partial_\tau \vec N|^2 + r_N |\vec N|^2 + u_N|\vec N|^4 \\
{\cal L}_{bN} & = v|b|^2 |\vec N|^2.
\end{align}
At the multicritical point where both $b$ and $\vec N$ are critical, a small coupling $v$ is an irrelevant perturbation\cite{aharony}, thus we are left with a decoupled $O(3) \times O(2)$ fixed point. As before, we now need to consider the effects of coupling to the Fermi surface and to the lattice. The irrelevance of the direct coupling between $b$ and the Fermi surface, as well as of the lattice pinning follow by the same argument as for pure charge-stripes melting transitions. Like in the case of charge stripes, there is a symmetry allowed coupling between the energy density of the $\vec N$ fluctuations and soft shape-fluctuations of the Fermi surface. Since $\nu_{O(3)} > \frac{2}{3}$, this is again irrelevant. Next, the field $\vec N$ is not physical (gauge invariant) and thus cannot directly couple to the Fermi surface. The gauge invariant spin-density that coupled to the spin density of the itinerant electrons at the Hot spots is $\vec M = b\vec N$. In real space and (imaginary) time the correlation function of $\vec M$ simply factorizes into the correlators of its constituents, i.e.
\begin{equation}
\langle \vec M({\bf r}, \tau)\cdot \vec M( 0, 0)\rangle \sim \frac{1}{\left({\bf r}^2 + v_c^2\tau^2\right)^{\frac{1+\eta_b}{2}}\left({\bf r}^2 + v_s^2\tau^2\right)^{\frac{1+\eta_s}{2}}}.\label{eqn:spincorr}
\end{equation}
In particular, we can read off the anomalous dimension of $\vec M$ as $\eta_M = 1 + \eta_N + \eta_b$. Here $\eta_{N,b}$ are the (small) anomalous dimensions of the $b$ and $\vec N$ fields in the regular 3D $XY$ and $O(3)$ models, respectively. In the presence of a Fermi surface, the fluctuations of the physical spin fluctuations $\vec M=b\vec N$ will be Landau damped, like the fluctuations of the physical charge fluctuations $\psi=b^2$ in Eqn. \ref{ldcs}, i.e.
\begin{equation}
\int d^2q d \omega |\omega| |\vec M|^2.
\end{equation}
Since $\eta_M>1$, the Landau damping of spin-fluctuations is also irrelevant. A further possibility to construct a gauge invariant operator as a combination of $b$ and $\vec N$ fields is the spin quadrupole operator $Q_{ab} = N_a N_b - \frac{1}{3} \vec N^2 \delta_{ab}$. It couples to the microscopic quadrupole operator of the conduction electrons, which is given by
\small
\begin{align}
Q^\text{el}_{\alpha \beta}(\vect r_1,\vect r_2) =f(\vect r_1 - \vect r_2) \(\frac{\sigma_1^\alpha \sigma_2^\beta +\sigma_1^\beta \sigma_2^\alpha}{2}-\frac{\delta^{\alpha \beta}\vec \sigma_1 \cdot \vec \sigma_2}{3} \),
\end{align}
\normalsize
where $\vec \sigma_i$ is the spin-density at $\vect r_i$ and $f(\vect r)$ is a non-universal function obeying lattice symmetries. For slow fluctuations of $Q^\text{el}_{\alpha \beta}$ all the electrons need to be on the Fermi-surface which imposes strong phase-space constraints familiar from Fermi-liquid theory. It is thus convenient to decompose the vertex into its angular harmonics in the particle-hole and Cooper channel, to get
\begin{align}
&\chi''_Q(\vect K,\vect \Omega)\\
 &=\int_0^\Omega d\omega\sum_{k}\sum_{m} F_m \Pi''_m(\vect k,\omega)\Pi''_{-m}(\vect K-\vect k,\Omega-\omega)\nonumber \\
&+\int_0^\Omega d\omega\sum_{k}\sum_{m} V_m C''_m(\vect k,\omega)C''_{-m}(\vect K-\vect k,\Omega-\omega)\nonumber,
\end{align}
 where $\Pi_m$ and $C_m$ are the particle-hole and particle-particle propagators, respectively, with angular momentum $m$. Now it is straightforward to evaluate $\chi_Q$ term by term and thus establish the irrelevance of Landau damping. In particular in the $m=0$ channel we have, for a Fermi-liquid
\begin{align}
&C_0''(\omega,k)\sim \Theta(\omega^2-v_F^2 k^2)\\
&\Pi_0''(\omega,k)\sim \frac{\omega}{\sqrt{(v_F k)^2-\omega^2}}\Theta((v_F k)^2 - \omega^2),
\end{align}
and it is then clear that $\chi''_Q \sim \Omega^3$ (with a cut-off dependent logarithm in the particle-hole channel) and hence damping is strongly irrelevant.

We have thus verified that the decoupled multicritical point is indeed stable both against coupling to the Fermi surface, and against pinning to the crystalline lattice. As we pointed out, this is the opposite of what one would expect for a conventional stripe melting transition, where the stripe fluctuations become strongly coupled to the Fermi surface at low energies. This immunity of the critical stripe fluctuations to the properties of the Fermi-surface, whether it is gapless or has a single particle (pseudo-)gap is consistent with the experiments of Ref. \onlinecite{AEPPLI}. More quantitatively, in the theory introduced above, stripe correlations at finite temperature obey $\omega/T$ scaling with a dynamical critical exponent $z=1$, as observed experimentally. Further, the height of the peak in the dynamical spin-susceptibility scales with temperature (see Sec. \ref{spinscaling}) as $\lim\limits_{\omega\rightarrow 0}\frac{\chi_P"(\omega, T)}{\omega}\sim \frac{1}{T^{3 - \eta_M}}$, where $\eta_M=1+\eta_N+\eta_b\approx 1.08$. This again agrees remarkably well with the results of Ref. \onlinecite{AEPPLI}.

\section{Single particle Green-function and Fermi surface reconstruction}\label{sec:single}
We now turn to calculate the Green-function of the electrons at the phase transition. Since the coupling to the fluctuating stripe-order is irrelevant, it is sufficient to consider the leading order contribution to the electron self-energy in perturbation theory. The same self-energy has already been evaluated explicitly in a different context in Ref. \onlinecite{tarun}. We here reproduce the universal behavior from scaling. Electrons on points of the Fermi-surface which are connected by $\vect Q$ (hot spots) will be affected most strongly, so we focus on those.
\begin{figure}[ht]
 \includegraphics[width=2.5cm]{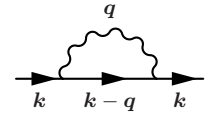}
\caption{Diagram for the leading contribution to the Fermion self-energy.}\label{fig:fse}
\end{figure}
In the vicinity of the critical point, the electron self-energy obeys (see Eq. (\ref{eqn:scaleselfenergy}))

\begin{align}
\Sigma\(\omega, \delta k, \kappa, \Delta_\text{stripe}\)=\omega g\(\frac{\delta k^z}{\omega},\frac{\Delta_\text{stripe}}{\omega}, \frac{\kappa_0}{\omega^{(1-\eta)/2}}\)\label{eqn:scaleselfenergy2},
\end{align}
where $\delta k$ is the momentum-deviation from the hot spot and $g$ is a universal scaling function. The leading order contribution (see FIG. \ref{fig:fse}) appears at order $\kappa^2$, so we get
\begin{align}
\Sigma\( \omega,\delta k =0, \kappa \rightarrow 0, \Delta_\text{stripe}=0\) \sim \kappa_0^2 \omega^\eta
\end{align}
 Since $\eta>1$ the Fermi-surface as well as the Landau quasiparticles remain sharply defined at this critical point, even at the hot-spots.

We can further use the scaling form of the self-energy to discuss the reconstruction of the Fermi surface. For sufficiently large doping $g>g_c$ the stripes are melted and we have a large Fermi surface. At lower doping the stripe fluctuations, and consequently the scattering of Fermions increases. It is clear that the scattering is non-uniform on the Fermi-surface, i.e. it decreases with distance to the hot-spot. At even lower doping, spectral weight from the original, large Fermi-surface starts being transferred to a ``shadow'' Fermi surface, i.e. the surface of $\vect k$ points that is connected to the large Fermi-surface by $\vect Q$. Deep in the stripe ordered phase, the shadow Fermi surface becomes part of the new, reconstructed Fermi surface. In the vicinity of the critical point, the spectral weight obeys universal scaling (with non-universal angle-dependent prefactors) which can be easily calculated.
\begin{figure}[ht]
 \includegraphics[width=8cm]{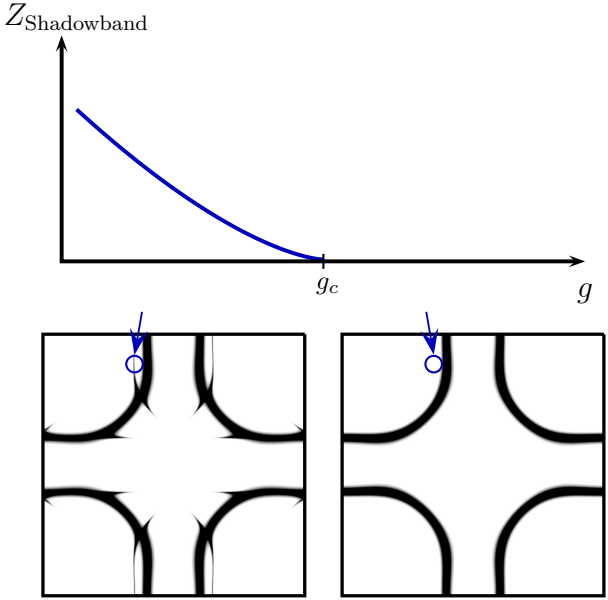}
\caption{The spectral weight on the shadowband vanishes as a power law close to the critical point.}\label{fig:shadow}
\end{figure}

We compute the self-energy for electrons (FIG. \ref{fig:fse}) close to a generic point $\vect K_0$ on this ``shadow" Fermi-surface, safely away from the hot spots. It is again given by Eq. (\ref{eqn:scaleselfenergy2}), since the intermediate electron lies on the Fermi surface. For $\omega/v_F, \delta k, \Delta_\text{stripe} /v_F\ll K_0$ the denominator in the imaginary part of the Green function
\begin{align}
 \text{Im} G(\omega, \vect K_0+\delta \vect k,\Delta_\text{stripe})= \frac{ \Sigma''}{\(\omega - \epsilon_{\vect K_0 + \delta \vect k }-{\Sigma'}\)^2+( {\Sigma''})^2}
\end{align}
can be replaced by $\epsilon_{\vect K_0}^2$ and the spectral function inherits the scaling form of $\Sigma''$, i.e.
\begin{align}
 A(\omega,\vect K_0 + \delta \vect k) \sim \frac{\omega^{\frac{\eta}{z}}}{\epsilon_{\vect K_0}^2} G(\frac{\delta k^z}{\omega},\frac{\Delta_\text{stripe}}{\omega}),
\end{align}
where $G$ is a new scaling function.

This implies \cite{criticalfs} that the quasiparticle weight $Z_\text{shadow}$ obeys a power law (see FIG. \ref{fig:shadow})
\begin{align}
Z_\text{shadow} \sim \frac{1}{\epsilon_{\vect K_0}^2}|g-g_c|^{\nu(z+\eta)}.
\end{align}
Here $\eta$ is the anomalous dimension of either charge of spin fluctuations, depending on which ordering is responsible for the reconstruction.

\section{Pairing by stripe fluctuations}\label{sec:pair}
We now turn to the possibility of superconductivity due to stripe fluctuations. We first focus on the stripe melted phase and the approach to the critical point. Since the coupling between the fermions on the Fermi-surface and the fluctuating stripes is irrelevant, it is safe to integrate out the stripe-fluctuations which gives rise to a 4-fermion interaction
\begin{align}
V(\omega,\vect k)= \frac{\kappa^2}{|\vect Q - \vect k|^{2-\eta_s}} W\(|\vect Q - \vect k|\xi,\omega \xi\),
\end{align}
where $W$ is a scaling function with.
\begin{align}
&\lim \limits_{x \rightarrow 0} W(x,0) = x^{2-\eta_s}
&\lim \limits_{x \rightarrow \infty} W(x,0) = 1
\end{align}
A sample function with these properties is
\begin{align}
\tilde W(x,0)= \frac{x^{2-\eta_s}}{x^{2-\eta_s}+1}
\end{align}
 This interaction can then be treated in mean-field theory.
\begin{figure}[ht]
\includegraphics[height=3.25cm]{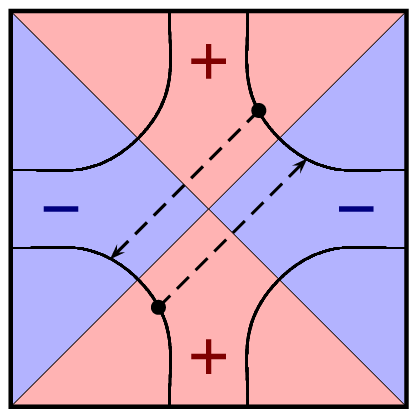}
\hspace{2mm}
 \includegraphics[height=3.25cm]{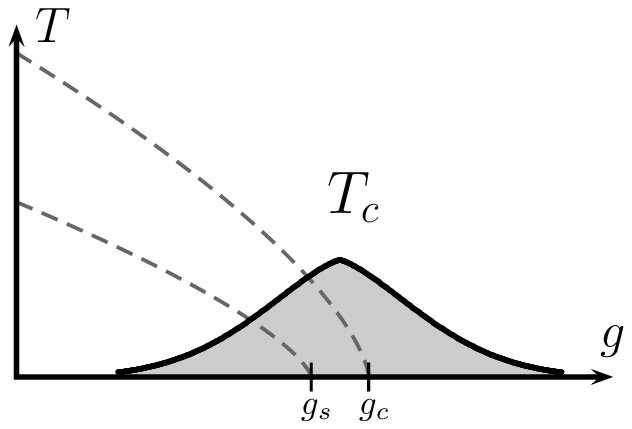}
\caption{(color online) Left: A typical large Fermi-surface of the cuprates is shown in the first Brillouin zone, where the shaded regions denote the sign of a $d_{x^2-y^2}$ order parameter. The solid dots denote a pair of electrons, which resides at momenta where the sign of the order parameter is positive. After scattering on a fluctuating stripe, the scattered Cooper pair, denoted by the tips of the arrows, find itself at momenta where the sign of the order parameter has changed. Since the interaction itself is repulsive, the effective electron-electron interaction is attractive in the $d_{x^2-y^2}$ channel . Right: Phase diagram including superconductivity induced by stripe fluctuations. The transition temperature $T_c$ is highest when spin fluctuations are critical.}\label{fig:supcon}
\end{figure}
 Since the interaction is repulsive, it can only contribute to a superconducting order parameter with opposite sign on hot-spots which are connected by $\vect Q_s$ (see FIG. \ref{fig:supcon}). For $\vect Q_s$ close to $(\pi,\pi)$ and a typical (large) Fermi-surface of the cuprates, the lowest angular momentum channel that satisfies this is\cite{dwave} $d_{x^2-y^2}$. To determine the mean-field gap at zero temperature, at weak coupling, we can neglect the frequency dependence of the interaction, i.e.
\begin{align}
&\Delta_d \sim \exp\[- \frac{1}{N(E_F) u_d}\]\\
&u_d= - \int d\vect k d\vect k' V(\vect k-\vect k')d_{\vect k}d_{\vect k'}\delta(\epsilon_{\vect k}-\mu)\delta(\epsilon_{\vect k'}-\mu),
\end{align}
where $d_{\vect k}$ is a function with $d_{x^2-y^2}$ symmetry. The dominant contributions to this integral come from momenta close to the hot-spots where $d_{\vect k}d_{\vect k'}\approx -1$. From these momenta we get
\begin{align}
 u_d(\xi,K)\approx \kappa ^2 \xi^{-\eta_s} U(\xi K),
\end{align}
where $K$ is a momentum cut-off and the scaling function $U$ must obeys
\begin{align}
&\lim \limits_{x \rightarrow \infty} U(x) = x^{\eta_s}
&\lim \limits_{x \rightarrow 0} U(x) = x^{2}.
\end{align}

Next we want to determine the transition temperature $T_c$. It is clear that for temperatures $T \gg \xi^{-1}$ the frequency dependence of the interaction can no longer be dropped. For a qualitative result it is sufficient to implement this approximately by writing the temperature dependent effective interaction as
\begin{align}
 u_d(\xi,K;T)\approx u_d(\sqrt{\xi^2+T^{-2}},K)
\end{align}
For weak coupling, the self-consistent equation for $T_c$ then becomes
\begin{align}
\log \frac{\Lambda}{T_c}=\frac{1}{u_d(T_c) N(E_F)},
\end{align}
where $\Lambda$ is a UV cutoff. 

In the stripe ordered phase, the region around the hot-spots becomes gapped. This will suppress the ability of the residual stripe flutuations to cause pairing.
 The resultant phase-diagram is shown schematically in FIG. \ref{fig:supcon}. As expected we find that pairing is strongest close to the critical point, where it is mediated by gapless spin fluctuations.

\section{Experiments}\label{sec:exps}
In this section we discuss our results for the quantum critical stripe fluctuations in the context of available data.

\subsection{Scaling of quantum critical spin fluctuations}\label{spinscaling}
We first briefly review the standard scaling assumptions for the full $\vect q$ and $\omega$ dependent dynamical spin susceptibility
associated with quantum critical spin fluctuations. The imaginary part can be expressed in terms of a universal scaling function $F$ as
\begin{equation}
\chi''(\vect q, \omega; T, \delta) = \frac{c_0}{|\vect q - \vect Q_s|^{2-\eta}} F\left(\frac{\omega}{c_1 |\vect q- \vect Q_s|^z}, \frac{\omega}{T}, \frac{\omega}{E_0 \delta^{\nu z}} \right).
\end{equation}
As before, $\vect Q_s$ is the incommensurate peak wave vector, $T$ is temperature and $\delta = |g-g_c|$ is the distance from the $T=0$ quantum critical point. $z$ is the dynamical critical exponent, $\nu$ is the correlation length exponent, and $\eta_s$ is the anomalous dimension of the spin order parameter. $c_0, c_1, E_0$ are non-universal numbers.

First, at $\omega, T\gg E_0 \delta^{\nu z}$ the system is in the ``quantum critical regime" and does not know that the ground state is not exactly at the quantum critical point. In this regime the width of the peak in $q$-space will satisfy
\begin{equation}
\label{kappawt}
\kappa(\omega, T) = \left(\frac{\omega}{c_1}\right)^{\frac{1}{z}} K\left(\frac{\omega}{T}\right)
\end{equation}
where $K$ is a universal scaling function.
Restricting to $\omega = 0$ then we get the well known result
\begin{equation}
\label{kappa0}
\kappa(\omega = 0, T\gg E_0 \delta^{\nu z}) \sim T^{\frac{1}{z}}.
\end{equation}

Second, exactly at the peak $\vect q = \vect Q$, we have (still in the ``quantum critical'' regime)
\begin{equation}
\label{chiP0}
\chi_P''(\omega; T) = \frac{a_o}{T^{\frac{2-\eta_s}{z}}} X\left(\frac{\omega}{T}\right)
\end{equation}
with $a_0 $ a non-universal number (related to $c_0, c_1$) and $X$ a universal scaling function.
At low frequency at a non-zero $T$, $\chi''$ is linear in $\omega$ so that the scaling function $X(x) \sim x$ for small $x$. This then gives
\begin{equation}
\label{chiP}
\lim_{\omega \rightarrow 0} \frac{\chi_P''(\omega, T\gg E_0 \delta^{\nu z})}{\omega} = \frac{\tilde{a}_0}{T^{(2-\eta_s+z)/z}}
\end{equation}
\subsection{Scaling in the LSCO experiment of Ref. \onlinecite{AEPPLI}}
We will now use the scaling properties reviewed above to carefully discuss the experiment in Ref. \onlinecite{AEPPLI}. There, Aeppli \emph{et. al.} used neutron scattering to find the spin structure factor $S(\vect Q,\omega;T)$ which is related to the imaginary part of the susceptibility via
\begin{align}
S(\vect Q,\omega;T) = \frac{\chi''(\vect Q,\omega;T)}{1-e^{-\frac{\omega}{T}}}\xrightarrow{\text{$\omega \ll T$}} \frac{\chi''(\vect Q,\omega;T) T}{\omega}.
\end{align}
In their data, they fit the temperature dependence of the peak height to a power-law (Fig. 3B in Ref. \onlinecite{AEPPLI} )
\begin{align}
\lim_{\omega \rightarrow 0} \frac{\chi_{P,\text{Fit}}''(\omega, T\gg E_0 \delta^{\nu z})}{\omega} \sim T^{-1.94},
\end{align}
from which it follows (\ref{chiP}) that
\begin{align}
\frac{2-\eta_s+z}{z} \approx 1.94.
\end{align}
To determine $z$, the width of the peak $\kappa$ is plotted as a function of $\sqrt{\omega^2+T^2}$ over a range of frequencies and temperatures (Fig. 4 in Ref. \onlinecite{AEPPLI} ). A linear relationship
\begin{align}
\kappa(\omega,T)\sim \sqrt{\omega^2+T^2}
\end{align}
is found, which implies (\ref{kappawt}) that $z=1$, which in turn determines the anomalous dimension $\eta_s \approx 1.06$.

Some caution is however required with the quantum critical interpretation of the data. To emphasize this we note that Ref. \onlinecite{AEPPLI} also plotted $\chi_P''(\omega,T)$ as a function of $\kappa(\omega,T)$ (Inset of Fig. 4 in Ref. \onlinecite{AEPPLI} ) and from a fit to the data found that
\begin{align}
\lim_{\omega \rightarrow 0} \frac{\chi_{P,\text{Fit}}''(\omega, T\gg E_0 \delta^{\nu z})}{\omega} \sim \kappa^{-3}(\omega=0,T).\label{kappa3}
\end{align}
If quantum critical scaling applies, then from Eq. (\ref{kappa0},\ref{chiP}) it follows that
\begin{align}
\lim_{\omega \rightarrow 0} \frac{\chi_{P}''(\omega, T\gg E_0 \delta^{\nu z})}{\omega} \sim \kappa^{\eta_s-3}(\omega=0,T).
\end{align}
Thus the last fit implies $\eta_s=0$, in apparent contradiction to the conclusion $\eta_s=1.06$, reached from the previous fits.

To address this question, it is necessary to analyze $\kappa(\omega,T)$, which relates the two apparently contradictory fits. $\kappa(\omega, T)$ saturates at low $T, \omega$. One possible cause is of course that the system is not exactly at the quantum critical point. A second possibility is quenched disorder. As the stripe order parameter breaks translation symmetry, non-magnetic disorder will couple to it roughly as a ``random field". This will always lead to a saturation of $\kappa$ - even when there is true stripe ordering.

One resolution of this apparent contradiction is that the fit leading to (\ref{kappa3}) includes data from the region where $\kappa$ has saturated. If the saturation is mainly due to disorder, those data points (small values of $\kappa$) should not be used in scaling plots. The remaining data points can be fit with
\begin{align}
\lim_{\omega \rightarrow 0} \frac{\chi_{P,\text{Fit}}''(\omega, T\gg E_0 \delta^{\nu z})}{\omega} \sim \kappa^{-2}(\omega=0,T)\label{kappa2},
\end{align}
in agreement with $\eta_s \approx 1.06$. Of course, once several data points are excluded, the range of the remaining data is rather small and consequently the error in extracting the exponent is large.

In a more recent paper\cite{walstedt}, the same neutron data is used to compute
\begin{equation}
\tau_\text{eff}=T\int_{\vect q} \lim \limits_{\omega \rightarrow 0}\frac{\chi"(\vect q, \omega, T)}{\omega} .
\end{equation}
If the $\vect q$ integral is restricted to the scaling region around $\vect Q$, scaling applies and the contribution of those wave-vectors to $\tau_\text{eff}$ is given by
\begin{equation}
\tau_\text{eff,scal} \equiv T \int_{\vect q \approx \vect Q} \lim \limits_{\omega \rightarrow 0}\frac{\chi"(\vect q, \omega, T)}{\omega} \sim T^{\eta_s}
\end{equation}
Thus, since $\eta_s$ is close to one, $\tau_\text{eff,scal}$ must have a nearly linear $T$ dependence. In the data of Ref. \onlinecite{walstedt}, $\tau_\text{eff}$ is roughly $T$-independent. If scaling is indeed satisfied by the neutron data near $\vect Q$ then it must be that the $q$-integral is dominated by non-scaling contributions away from the peak.

These caveats about the evidence for critical scaling in the existing data strongly emphasizes the need for further experimental efforts to study the singular stripe fluctuations in modern samples of cuprates. Little is also known about the doping and magnetic field dependence of these singular fluctuations. In light of the suggested crucial role\cite{taillefer} of these fluctuations for the physics of the strange metal and the theory discussed in this paper we hope that such experimental efforts will be forthcoming in the near future.

\subsection{Other scaling in cuprate neutron experiments}
Scaling of the spin fluctuation spectrum has been reported in a number of cuprates over the years\cite{keimer2,stock1,stock2}. In an early paper, Keimer et al\cite{keimer2} studied very lightly doped LSCO and found evidence of scaling of the local ({\em i.e} $q$-integrated) dynamic susceptibility as a function of the ratio $\omega/T$. Subsequent experiments on highly underdoped YBCO near the edge of the superconducting boundary also saw similar $\omega/T$ scaling\cite{stock1}. This scaling in very underdoped cuprate samples presumably has little to do with the stripe quantum criticality discussed in this paper. 

More pertinent possibly is the $\omega/T$ scaling seen\cite{stock2} in the normal state of underdoped YBCO$_{6.5}$. This is in the pseudogap regime and the spin fluctuation spectrum has incommensurate peaks (``dynamic stripes"). Detailed studies of the $q$ and $T$ dependence of the scale invariant part of the scattering have not been reported in the literature. In the context of this paper it is particularly interesting to ask if this scaling is associated with the `mother' multicritical regime (see Fig. \ref{fig:sn1}) at intermediate temperatures. If so then the detailed $q$, $\omega$ and $T$ dependence of $\chi"$ will satisfy the scaling form discussed above with the exponent $\eta_s \approx 1.06, z = 1$. Upon cooling superconductivity develops and the measured spin fluctuations change character. It is interesting to ask what happens if the superconductivity is suppressed in a field. At similar doping levels in YBCO recent NMR experiments\cite{julien} show that there is charge stripe order without any accompanying spin stripe order. In terms of Fig. \ref{fig:sn1} this places the system between $g_c$ and $g_s$. In this regime in our theory there are dynamic scale invariant stripe fluctuations at intermediate temperatures. Upon cooling the spin correlation length saturates to a finite value but there is a charge ordering transition. Associated with this there is a reconstruction of the Fermi surface.

\subsection{Dynamic susceptibility}
\begin{figure}
\includegraphics[width=\columnwidth]{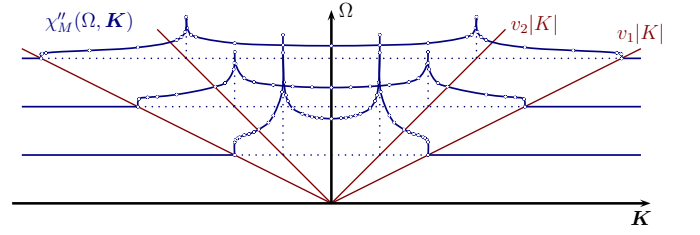}
\caption{The imaginary part of the spin-susceptibility, plotted for three fixed values of $\Omega$, as a function of $K$ (MDC). The value of $\chi''(\Omega,\vect K)$ is given by the height of the blue curve above the (dotted) base line. For any given value $\Omega$ it is non-zero only over a finite interval. For fixed $K$ and as a function of $\Omega$ (EDC) it has a similarly sharp onset at $v_1 K$, but does not drop back to zero at higher frequencies.}
\label{fig:mdc}
\end{figure}

For comparison with future experiments, we now compute the dynamic spin-susceptibility at the multicritical point, i.e. the Fourier-transform of Eq. (\ref{eqn:spincorr}):\small
\begin{align}
&\chi_M''(\Omega,\vect K)\sim\\
&\int_0^\Omega d\omega \int d^2 k \frac{\Theta(\omega^2 - v_1^2 k^2)}{(\omega^2 - v_1^2 k^2)^{\frac{2-\eta_1}{2}}}
\frac{\Theta((\Omega-\omega)^2 - v_2^2 (\vect k-\vect K)^2)}{((\Omega-\omega)^2 - v_2^2 (\vect k-\vect K)^2)^{\frac{2-\eta_2}{2}}} \nonumber
\end{align}
\normalsize
Two trivial limits of this function are given by
\begin{align}
&\chi_M''(\Omega,0)\sim \Omega^{\eta_1+\eta_2-1}\\
& \chi_M''(\Omega,\vect K)\big|_{v_1=v_2}\sim \frac{\Theta(\Omega^2 - v_1^2 K^2)}{(\Omega^2 - v_1^2 K^2)^{\frac{1-\eta_1-\eta_2}{2}}}.
\end{align}
For finite $\vect K$ and $v_1 \neq v_2$ one can see that the most singular contributions to $\chi_M''(\Omega,\vect K)$ occur when \emph{all} the momentum and frequency are carried by either $b$ or $\vec N$, which is possible only for $\Omega = v_1 K$ or $\Omega = v_2 K$. For example if $\Omega = v_2 K$ then the integrand diverges at $\omega,k \rightarrow 0$ as $\omega^{\frac{\eta_2}{2}+\eta_1-3}$. After integration a power-law singularity
\begin{align}
 \chi_M''(v_2 K+\delta\Omega,\vect K)-\chi_M''(v_2 K,\vect K)\sim \delta\Omega^{\frac{\eta_2}{2}+\eta_1}
\end{align}
remains. A numerical plot of this function is shown in FIG. \ref{fig:mdc}. In any actual experiment, $\chi_M$ will be modified from this by finite temperature effects. In particular, all sharp features (onset and peaks) are expected to be washed out. We expect that what remains as a qualitative feature is a dispersive peak which is relatively narrow as a function of $\vect K$ but has a long tail as function of $\Omega$.

\subsection{Scaling of quantum critical charge fluctuations}
In addition to measuring critical spin fluctuations via neutron scattering, it may be possible to detect critical charge fluctuations, e.g. via X-ray experiments. At a quantum critical point where charge order melts, the scaling properties of the dynamical charge structure factor $S_\text{Charge}(\vect Q,\Omega;T)$ follow from the same analysis as in Sec. \ref{spinscaling}. In particular, right at the critical point, the peak amplitude at low frequencies obeys
\begin{align}
S_{P,\text{Charge}}(T)\sim T^{(\eta_c-2)/z},
\end{align}
where $\eta_c$ is the anomalous exponent of charge fluctuations, which is $\eta_c \approx 1.49$ in our theory. Measuring this exponent thus constitutes an independent test of our theory.

\subsection{Tunneling density of states near pinned stripes}
STM has been a highly successful tool in investigation of striped phases in the cuprates. Recently, it has been used to image individual defects in the stripe order of BSCCO\cite{stmvortex}. Within our theory as the stripe melting transition is approached the core energy of doubled dislocations decreases to zero while that of single ones stays non-zero. It is
possible therefore that upon approaching optimal doping the density of close dislocation pairs increases at the expense of isolated dislocations. It will be interesting to explore this in STM
experiments.

STM experiments can also potentially be used to determine the anomalous dimension of the charge stripe fluctuations $\eta_c$. In the stripe-melted phase, but close to the phase transition, there will always be disorder which locally pinns fluctuating stripes. Close to such an impurity, there is then local stripe-order with an amplitude which decays with distance from the impurity. By locally measuring the tunneling density of states, these oscillations, as well as their decay, can be detected experimentally.

\begin{figure}[ht]
\includegraphics[width=3cm]{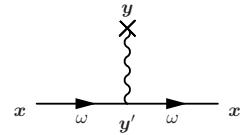}
\caption{The lowest order contribution of pinned stripes to the electronic tunneling density of states. The box denotes a time-independent disorder potential}
\label{fig:stm}
\end{figure}

\begin{figure}[ht]
\includegraphics[width=8cm]{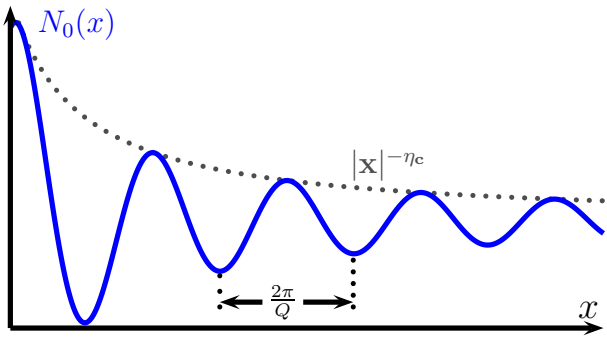}
\caption{Density of states for tunneling in electrons at zero bias as a function of distance from an impurity. The impurity locally pins stripe fluctuations which lead to an oscillatory charge density close to the impurity. Conduction electrons scatter on this pinned stripe fluctuations which leads to an oscillatory tunneling density of states.}
\label{fig:stmplot}
\end{figure}

The lowest order oscillatory contribution to the tunneling density of states is shown in FIG. \ref{fig:stm}. The contribution due a delta-function disorder potential at the origin is
\begin{align}
N_{\omega}(\vect x)\sim \int_{\vect y} G_{\vect x - \vect y}(\omega)\cos \(\vect Q \cdot \vect y\)D_{\vect y}(\Omega =0) G_{\vect y - \vect x}(\omega)
\end{align}

 We note that the integral is dominated by electrons close points on the contour of energy $\mu + \omega$ which are connected by $\vect Q$. Around these points we can expand the dispersion to linear order, i.e. under the integral we have $G_\omega(x)\sim \frac{1}{x}$. The propagator of charge-stripe fluctuations at zero frequency is $D(x) \sim \frac{1}{x^{\eta_c}}$ so we get
\begin{align}
N_{\omega=0}(\vect x)\sim |\vect x|^{-\eta_c}\cos \(\vect Q \cdot \vect x+\Phi\),
\end{align}
where $\Phi$ is some constant. The spatial decay in the amplitude of the quasiparticle oscillations is directly given by the anomalous exponent $\eta_c$ of the charge stripes (see FIG. \ref{fig:stm}). We note that unlike quasiparticle interference, these oscillations are \emph{non-dispersive}, i.e. the wave-vector $\vect Q$ is set by the stripe order, and does not change as a function of bias $\omega$.

\section{Conclusions}
In this paper we have studied several examples of continuous quantum melting phase transitions of stripe order motivated by the cuprates. The stripe melting was assumed to be driven by the proliferation of doubled dislocations. We developed a simple and intuitive picture of the resulting stripe melted phase: the fluctuating stripes form closed loop configurations of arbitrary size. Cutting a loop open to produce two end points costs finite energy. We therefore dubbed this phase a ``Stripe Loop Metal". A Stripe Loop Metal phase has sharply defined Landau quasiparticles at a Fermi surface that satisfies the usual Luttinger theorem. It however differs from the conventional Fermi liquid in some very subtle ways. The spectrum of excitations of the stripe order parameter is fractionalized. Associated with this there is a gapped $Z_2$ topological defect that is the end point of an open stripe(equivalently a relic of the single stripe dislocation of the ordered phase). These differences are sufficiently delicate that even if such a stripe melted state exists in a material, it cannot easily be distinguished from a conventional Landau Fermi liquid.

We showed that this kind of stripe melting transition is strongly coupled yet tractable. The critical stripe fluctuations dynamically decouple from the Fermi surface. However, in the ordered phase, the stripes recouple to the Fermi surface and reconstruct it. Thus our results provide concrete examples of tractable strongly coupled quantum phase transitions associated with stripe melting and the associated reconstruction of the Fermi surface. We are not aware of any other controlled theories of quantum criticality related to Fermi surface reconstruction by translation symmetry breaking order in two space dimensions.

In the examples studied in this paper the dangerous irrelevance of the coupling of the stripe fluctuations to the Fermi surface implies that the latter reconstructs at a scale that is parametrically lower than the scale at which the stripe ordering occurs. Furthermore the stripe fluctuations do not destroy the Landau quasiparticle at any point of the Fermi surface at the quantum critical point.

Before concluding we discuss these results in the context of several recent experiments and theoretical suggestions on the cuprates. As discussed in the introduction there is growing evidence for the ubiquity of stripe order in the underdoped cuprates. Stripe order has also been invoked to reconstruct a large Fermi surface to obtain pockets which may explain quantum oscillation phenomena in a magnetic field at low-$T$, and Hall effect and other measurements. In contrast there is very little evidence for stripe ordering in the overdoped side. A natural suggestion is that a quantum critical point associated with melting of stripe order and the related Fermi surface change underlies some of the strange normal state properties around optimal doping.

To apply the results of this paper to the cuprates we first need to postulate that the overdoped metal is a Stripe Loop Metal. There is very little experimental evidence against such a proposal, and so it is worthwhile to examine our results further. As proposed in our earlier work\cite{short} and discussed further in this paper, the
`mother' multicritical fixed point where the charge and spin stripe orders simultaneously quantum melt into an overdoped Stripe Loop Metal might potentially control the physics of the stripe fluctuations around optimal doping in the strange metal regime. This proposal finds some support in well known early experiments of Aeppli et al\cite{AEPPLI} on near-optimal La$_{2-x}$Sr$_x$CuO$_4$ providing an explanation of both the scaling of the peak width and peak height as a function of temperature/frequency. We outlined a number of other predictions from the proposed theory for future experiments.

A disappointing aspect of our results is that the critical stripe fluctuations associated with the transition to the Stripe Loop Metal are not capable of producing most of the observed non-Fermi liquid physics of the strange metal regime. In particular the electron quasiparticle remains well defined even at the stripe melting quantum critical point. Given the success of our theory in describing the observed stripe fluctuations in neutron data one possibility is simply that the physics driving the strange metal non-fermi liquid is not stripe quantum criticality (contrary to the proposal of Ref. \onlinecite{taillefer}). We suggest that our theory correctly describes the stripe sector which decouples dynamically from the electronic fluctuations of a non-fermi liquid metal at the stripe quantum critical point. The explanation of the non-Fermi liquid physics itself should then be sought in
some other mechanism.

We thank G. Aeppli, J. C. Davis, Eduardo Fradkin, Eun-Ah Kim, Steve Kivelson, Michael Lawler, Young Lee and Stephen Hayden for useful discussions. TS was supported by NSF Grant DMR-1005434.

\appendix
\section{Landau-damping in some non-Fermi liquid metals}\label{app:damping}

Consider a metal with a sharp Fermi surface, possibly in a Non-Fermi Liquid phase. We take a scaling form for the electronic spectral function
\begin{align}
A(\delta k,\omega)=\frac{1}{\omega^{\alpha/z_f}}F(\frac{\omega}{\delta k^{z_f}}),
\end{align}
where $\delta k = (\vect k -\vect k_F)\cdot \hat k_F$. This is appropriate for the Fermi-Liquid ($\alpha = z_f=1$) but also for the marginal Fermi-Liquid and certain non-Fermi Liquids, such as a metal at a Pomeranchuk transition ($\alpha=1,\ z_f=3/2$). Using this to perturbatively calculate the self-energy of the fluctuations of the stripe order parameter (FIG. \ref{fig:bosonselfenergy}), we find
\begin{align}
\Pi''(\omega,\vect Q)&= \int_0^\omega d\Omega \int d^2k A(\vect k, \omega+\Omega)A(\vect k, \omega)\\
&\sim \omega^{1+2\frac{1-\alpha}{z_f}}\label{eqn:landaudamp}.
\end{align}
At low energies, for $\alpha>\frac{2-z_f}{2}$, this dominates over the bare $\omega^2$ term and consequently $z$ is renormalized away from one.

 This remains true even for some more exotic models of the strange metal, such as the uniform Resonating Valence Bond (RVB) state studied, e.g., in Ref. \onlinecite{leenagaosa}.
In this model there is spin-charge separation, with the spin carried by charge-neutral fermions (spinons) which form a Fermi surface, and the charge carried by spinless bosons (holons) which form an incoherent Boltzmann gas. The electron Green function is given by the convolution of a spinon and a holon Green function. Nevertheless, at finite wave-vectors the Landau-damping again takes the form of Eq. (\ref{eqn:landaudamp}). The spin response of this state is determined solely by the spinons (which form a Fermi surface), thus the self-energy of spin stripe fluctuations has the form of Eq. (\ref{eqn:landaudamp}). The charge response is given by the Ioffe-Larkin rule
\begin{align}
\Pi_\text{electron} = \frac{\Pi_\text{spinon}\Pi_\text{holon}}{\Pi_\text{spinon}+\Pi_\text{holon}},
\end{align}
where the holon-polarizability is finite and non-singular at finite wave-vectors and $\Pi''_\text{spinon}$ is of the form of Eq. (\ref{eqn:landaudamp}). Expanding $\Pi''_\text{electron}$ for small $\omega$ then yields $\Pi''_\text{electron}\sim \omega^{1+2\frac{1-\alpha}{z_f}}$.

\section{RG-equations for the Tetragonal model}
Consider the theory described by Eq. (\ref{tetraaction}), i.e.
\small
\begin{align}
S=& \sum_{i=x,y}\int_{\tau,\vect r} (\partial_\mu \vec\psi_i)^2 + r (\vec\psi_i)^2+u(\vec\psi_i)^4+2v \int_{\tau,\vect r} (\vec\psi_x)^2 (\vec\psi_y)^2,
\end{align}
\normalsize
where we wrote $\vec \psi_i = (\text{Re}[b_i],\text{Im}[b_i])$. More generally we can consider this theory for $N$-component vectors $\vec\psi_i$. The RG-equations to leading order in $\epsilon = 4-d$ are readily obtained
\begin{align}
 &\frac{d}{dl} u =\epsilon u- (8+N) u^2 I_0- N v^2 I_0\\
&\frac{d}{dl} v =\epsilon v- (4+2N)uv I_0+4 v^2 I_0,
\end{align}
where $I_0$ depends on the cut-off scheme. The fixed points(FPs) are\small
\begin{align}
A: \ \ \ (u^*,v^*)=&(0,0)\\
B: \ \ \ (u^*,v^*)=&\(\frac{1}{8+N},0\)\epsilon/I_0\\
C: \ \ \ (u^*,v^*)=&\(\frac{N}{16+2N^2},\frac{N-4}{16+2N^2}\)\epsilon/I_0\\
D: \ \ \ (u^*,v^*)=&\(\frac{1}{8+2N},\frac{1}{8+2N}\)\epsilon/I_0.
\end{align}\normalsize
A standard stability analysis finds that the $O(2N)$-fixed point $D$ is stable for $N<2$, $C$ is stable for $2<N<4$ and the decoupled FP $B$ is stable for $N>4$. However it is known from higher-order expansions and Monte-Carlo studies that for $N>1$ the decoupled FP is the only stable fixed point in $d=3$. So the leading order $\epsilon$-expansion gives the incorrect result. Similarly, the leading order expansion in $1/N$ finds the decoupled FP to be unstable (this can be seen from Eq.(\ref{vscaling}), where to leading order in $1/N$, $\nu=1/2$). A higher order expansion in the presence of the nematic coupling (\ref{nemterm}), e.g. along the lines of Ref. \onlinecite{subir}, is rather involved, and we do not pursue this direction here.

\section{Dual description of the deconfined transition}\label{app:dual}
First consider the orthorhombic case. The dual description of the usual $XY$-transition is given by a vortex field $\Psi(\vect r) = e^{i \varphi(\vect r)}$ which is minimally coupled to a fluctuating $U(1)$ gauge field, i.e. the hopping of vortices is described by
\begin{align}
H_v = - t_v \sum_{\vect{r},\hat e} \exp \{i \varphi({\vect r}) - i \varphi({\vect r + \hat e})- 2\pi i a^{\hat e}({\vect r})\}+ h.c.,
\end{align}
where the number of the original $XY$-bosons is related to flux of the gauge field via
\begin{align}
 n_{\vect r} = a^{\hat y}(\vect r+\hat x)-a^{\hat y}({\vect r}) -a^{\hat x}({\vect r+\hat y})+a^{\hat x}({\vect r})= \vect \nabla \times \vect a.
\end{align}
In the following we will use $\vect \nabla$ to describe lattice derivatives to lighten the notation. When the vortices are condensed $\langle \Psi \rangle \neq 0$, single valuedness of $\Psi$ requires that
\begin{align}
 2\pi \sum_{\ell \in \Box}\hat \ell \cdot \vect a = 2\pi \vect \nabla \times \vect a=2\pi N,
\end{align}
i.e. the boson number is quantized in integers. For the $XY^*$ transition, we only allow doubled vortices $\Phi(\vect r) = e^{2 i\varphi(\vect r)}$. Their hopping is described by
\begin{align}
H^*_v = - t_v \sum_{\vect{r}}\exp \{i 2 \vect \nabla \varphi({\vect r}) - 4\pi i \vect a({\vect r})\}+ h.c.,
\end{align}
and when $\langle \Phi \rangle \neq 0$, single valuedness of $\Phi$ implies
\begin{align}
4\pi \vect \nabla \times \vect a=2\pi N,
\end{align}
i.e. the boson number is quantized in half-integers, in agreement with our fractionalizion prescription
\begin{align}
 \psi = b^2.
\end{align}
Further, the single-vortex is now an Ising variable, i.e. $\Psi = \pm \sqrt{\langle \Phi\rangle}$ and can be identified with the $Z_2$ gauge field. \\
Now we turn to the tetragonal case, where we have two vortex fields $\Psi_{i}(\vect r)=e^{i \varphi_{i}(\vect r)}$, each coupled to their own gauge-field $a_{i}(\vect r)$. We only allow hopping of pairs $\Psi_{1}\Psi_{2}$ or $\Psi_{1}\Psi_{2}^*$, i.e.
\begin{align}
H^\text{tet}_v = - t_v \sum_{\vect{r},\sigma=\pm}\exp &\big \{i \vect \nabla \varphi_1({\vect r})+\sigma i \vect \nabla \varphi_2({\vect r})\\
& - 2\pi i (\vect a_1({\vect r})+\sigma \vect a_2({\vect r}))\big\}+ h.c.\nonumber.
\end{align}
If $\langle\Psi_{1}\Psi_{2}\rangle=\langle\Psi_{1}\Psi_{2}^*\rangle\neq 0$ the fluxes are quantized as
\begin{align}
\vect \nabla \times\vect a_1 \pm \vect \nabla \times\vect a_2 = N_\pm,
\end{align}
or alternatively the boson numbers as
\begin{align}
 & n_1=\frac{1}{2}(N_++N_-)\\
& n_2 = \frac{1}{2}(N_+-N_-),
\end{align}
so in terms of the charges $Q_1,Q_2$ of the original $XY$-fields, the elementary excitations in the phase where the paired vortices have condensed carry charge $(Q_1,Q_2) = (1/2,\pm 1/2)$, which is reflected in the fractionalization prescription (Eq. (\ref{eqn:tetrafrac}))
\begin{align}
&\psi_x = b_1 b_2\nonumber \\
&\psi_y = b_1 b_2^*.
\end{align}

\end{document}